\documentclass[11pt,preprint]{aastex}
\usepackage{apjfonts}

\def\gax{\mathrel{\raise.3ex\hbox{$>$}\mkern-14mu\lower0.6ex\hbox{$\sim$}}}
\def\lax{\mathrel{\raise.3ex\hbox{$<$}\mkern-14mu\lower0.6ex\hbox{$\sim$}}}
\def\gtorder{\mathrel{\raise.3ex\hbox{$>$}\mkern-14mu
             \lower0.6ex\hbox{$\sim$}}}
\def\ltorder{\mathrel{\raise.3ex\hbox{$<$}\mkern-14mu
             \lower0.6ex\hbox{$\sim$}}}

\begin{document}

\title{Dusty Explosions from Dusty Progenitors: \\
     The Physics of SN~2008S and the 2008 NGC~300-OT}

\author{
  C.~S. Kochanek$^{1,2}$, 
  }

\altaffiltext{1}{Department of Astronomy, The Ohio State University, 140 West 18th Avenue, Columbus OH 43210}
\altaffiltext{2}{Center for Cosmology and AstroParticle Physics, The Ohio State University, 191 W. Woodruff Avenue, Columbus OH 43210}

\begin{abstract}
SN~2008S and the 2008 NGC~300-OT were explosive transients of stars self-obscured
by very dense, dusty stellar winds.  An explosive transient with an un-observed 
shock break-out luminosity of order $10^{10}L_\odot$ is required to render the 
transients little obscured and visible in the optical at their peaks. Such a
large break-out luminosity then implies that 
the progenitor stars were cool, red supergiants, most probably $\sim 9 M_\odot$ 
extreme AGB (EAGB) stars.  As the shocks generated by the explosions propagate outward through the dense wind,
they produce a shock luminosity in soft X-rays that powers the long-lived luminosity
of the transients.  Unlike typical cases of transients exploding into 
a surrounding circumstellar medium, the progenitor winds in these systems are optically
thick to soft X-rays, easily absorb radio emission and rapidly reform dust destroyed
by the peak luminosity of the transients.  As a result, X-rays are absorbed by the gas and 
the energy is ultimately radiated by the reformed dust.  Three years post-peak, both
systems are still significantly more luminous than their progenitor stars, but they 
are again fully shrouded by the re-formed dust and only visible in the mid-IR.  The high
luminosity and heavy obscuration may make it difficult to determine the survival
of the progenitor stars for $\sim 10$~years.  However, our model indicates that
SN~2008S, but not the NGC~300-OT, should now be a detectable X-ray source. 
SN~2008S has a higher estimated shock velocity and a lower density wind, so
the X-rays begin to escape at a much earlier phase.
\end{abstract}


\section{Introduction}
\label{sec:introduction}

SN~2008S (\citealt{Arbour2008}) and the 2008 NGC~300-OT (``optical transient'', \citealt{Monard2008}) were
initially classified as unremarkable Type~IIn transients that were likely massive 
star eruptions rather than true supernovae (\citealt{Steele2008}, \citealt{Bond2008}). 
However, searches for optical progenitors proved fruitless (\citealt{Prieto2008},
\citealt{Berger2008}), which was peculiar because the transients were little obscured
and a massive star should have been easily visible.  It became clear that they
represented a new class of massive star transients when \cite{Prieto2008a}
and \cite{Prieto2008} identified the progenitors in archival Spitzer data as
completely obscured stars with luminosities of order $60000L_\odot$ and apparent
photospheric temperatures of order $400$~K.  In mid-IR color-magnitude diagrams,
the stars lie at the extreme limit of the asymptotic branch (AGB)
and they are extremely rare, with only 1--10 similar stars per galaxy (\citealt{Thompson2009},
\citealt{Khan2010}). \cite{Thompson2009} suggested that the M85-OT-2006, SN~1999bw, 
and SN~2002bu are also likely members of this class, and \cite{Kasliwal2011} propose PTF10FQS
as a member.  They are clearly not members of the ``luminous red nova'' population
represented by V838~Mon and V4332~Sgr (\citealt{Kulkarni2007}).  
Our working hypothesis, which is by no means 
universally accepted, is that the progenitors are extreme AGB (EAGB) stars with masses of order 
$M_* \simeq 9M_\odot$ cloaked by extremely dense $\dot{M} \sim 10^{-4} M_\odot$/year winds.

While a few studies have focused on ``traditional'' optical examinations of these
transients (\citealt{Berger2009}, \citealt{Smith2009}, \citealt{Smith2011}), it is impossible to 
understand even their energetics without the addition of near-IR and mid-IR
data (\citealt{Bond2009}, \citealt{Botticella2009}, \citealt{Prieto2009},
\citealt{Ohsawa2010}, \citealt{Prieto2010a}, \citealt{Wesson2010}, 
\citealt{Hoffman2011}, \citealt{Prieto2011}).  Not only is the near/mid-IR
emission significant at peak, but it increasingly dominates the emission
as time passes, to the point where both transients are again invisible in the 
optical and near-IR but remain significantly more luminous than their
progenitors in the mid-IR (\citealt{Prieto2010a},
\citealt{Hoffman2011}, \citealt{Prieto2011}).  Thus, one goal of the 
present effort is to provide a self-consistent model of the evolving
spectral energy distributions (SED) of these transients. 

Unfortunately, there are two possible masses that can be associated with evolved
stars having the luminosities of the progenitors.  They can be cold (3000~K)
EAGB stars with $M_* \simeq 9 M_\odot$ or hotter ($>7500$~K) $\sim 15M_\odot$ stars, 
but we have no direct information on the progenitor temperatures because of the 
obscuration.  In \cite{Thompson2009}
we argued for the lower mass scale because extreme AGB stars are expected to
be heavily obscured by dusty winds while the more massive stars are not (e.g. \citealt{Poelarends2008}). 
This is supported by the lack of silicate features in the SEDs (e.g. \citealt{Wesson2010})
or in the IRS spectrum of the NGC~300-OT (\citealt{Prieto2009}), since
EAGB stars generally have graphitic dusts while massive stars have 
silicate dusts (e.g. \citealt{Groenewegen2009}).  \cite{Umana2010}, further cited by \cite{Smith2011},
 note the presence of PAH-related features
in mid-IR spectra of a candidate Luminous Blue Variable (LBV), and thus
argue that some more massive stars may also have graphitic dusts.  However,
\cite{Prieto2009} already noted that some massive stars have graphitic
dust features, but none unite the key points about the NGC~300-OT spectrum: the absence of silicate 
features, the presence of graphitic features and that the wavelengths of the
graphitic features are indicative of predominantly aliphatic hydrocarbons 
rather than PAHs, which implies formation in an environment with 
little UV radiation. This favors the EAGB model independent of the 
composition.  The highest mass stars
observed near the transients can be used to estimate an upper, {\it but not
a lower}, bound on the masses of the stars,  and there are stars
as massive as $15$-$20M_\odot$ near the NGC~300-OT (\citealt{Gogarten2009})
but not near SN~2008S (\citealt{Prieto2011}).  Because this approach
supplies only upper bounds, these results again favor the lower mass
EAGB stars if the two transients have a common physical origin.  Proponents of
the higher mass progenitors (\citealt{Bond2009}, \citealt{Smith2009}, 
\citealt{Berger2009}), also favor models in which the progenitor is 
obscured by a shell of previously ejected material, similar to
IRC$+$10420 (e.g. \citealt{Humphreys1997}), rather than a steady state wind.
The lack of mid-IR variability from
the progenitors (\citealt{Prieto2008}, \citealt{Thompson2009}) largely
ruled out absorption in an expanding shell since the expansion should
have led to an observable drop in the dust temperature over the duration
of the archival observations of the progenitors, again favoring the dense
winds observed around EAGB stars.  We will find that the
nature of the transient both constrains the radius of the progenitor
star to be very large and requires the existence of a very dense wind, 
further reinforcing the EAGB hypothesis.    

The nature of the transients is also debated.  In \cite{Thompson2009} we outlined 
a broad range of possibilities including stellar eruptions, sub-luminous SN 
and the final phases of stripping stars to leave a massive white dwarf.  Potentially
the most interesting of these possibilities is an electron capture SN (ecSN), since
they are predicted to be sub-luminous and associated with EAGB stars 
(e.g. \citealt{Pumo2009}).  \cite{Bond2009}, \cite{Berger2009}
and \cite{Smith2009} firmly favor non-explosive, stellar eruptions, \cite{Botticella2009}
favored the ecSN scenario and \cite{Kashi2010} suggested binary interactions. 
At present, the only evidence against explosive transients is the failure
to detect either X-ray or radio emission (\citealt{Chandra2008}, \citealt{Botticella2009}, \citealt{Berger2009}). 
The present day obscuration and luminosities mean that the simplest test of these
hypothesis, the survival of the progenitor, cannot be carried out.  While we will
be unable to answer this question precisely, we will find the transients had to
be explosive in nature.

Our goal here is to combine the available observations of the progenitors and the transients
to produce a self-consistent model that addresses the
puzzles about these events outlined above.  We make only three hypotheses.
First, the progenitors were obscured by dense dusty winds, whose properties we 
constrain in \S\ref{sec:progenitors}.  Second, that the transient ejected a significant
amount of mass at velocities characterized by the $500$-$1000$~km/s velocities 
observed near the transient peak (\citealt{Botticella2009}, \citealt{Bond2009}, \citealt{Smith2009}
\citealt{Berger2009}, \citealt{Smith2011}).  Third, that the dust 
destroyed by the transient subsequently reforms.  Section \S\ref{sec:transients} combines these 
elements to build quantitative models of the transient light curves in terms of the evolution
of the luminosity, shock radius, characteristic dust radius, and dust optical depth.
In \S\ref{sec:consequences} we examine the physical consequences of the model.
In \S\ref{sec:destroy} we show how the destruction of dust needed to render the
transients optically visible at peak requires an explosive transient from a cool, red 
supergiant. In \S\ref{sec:reform} we discuss recombination, cooling and dust re-formation
in the wind exterior to the expanding shock. In \S\ref{sec:xray} we propose that
the late time luminosity is powered by X-ray emission from shock heating
the dense wind that is subsequently absorbed and finally escapes in the
mid-IR. We also predict the X-ray emissions, finding that
while neither source should have been detected in existing X-ray observations,
SN~2008S should now be a detectable X-ray source while the NGC~300-OT should not.
Similarly, we expect early phase radio emission to be absorbed.  
Finally, in \S\ref{sec:optir} we discuss the expected evolution of the optical, near-IR
and mid-IR emissions and in \S\ref{sec:survivor} the possibilities for detecting surviving progenitor stars.
In \S\ref{sec:conclusions} we summarize the results, relate them to other 
similar transients and discuss future tests of the hypothesis.  We adopt distances
of $1.88$~Mpc for NGC~300 (\citealt{Gieren2005}) and $5.6$~Mpc for SN~2008S in NGC~6946
(\citealt{Sahu2006}) and Galactic foreground extinctions of $E(B-V)=0.013$ and $0.342$
based on \cite{Schlegel1998}.

\begin{figure}[t]
\centerline{\includegraphics[width=5.5in]{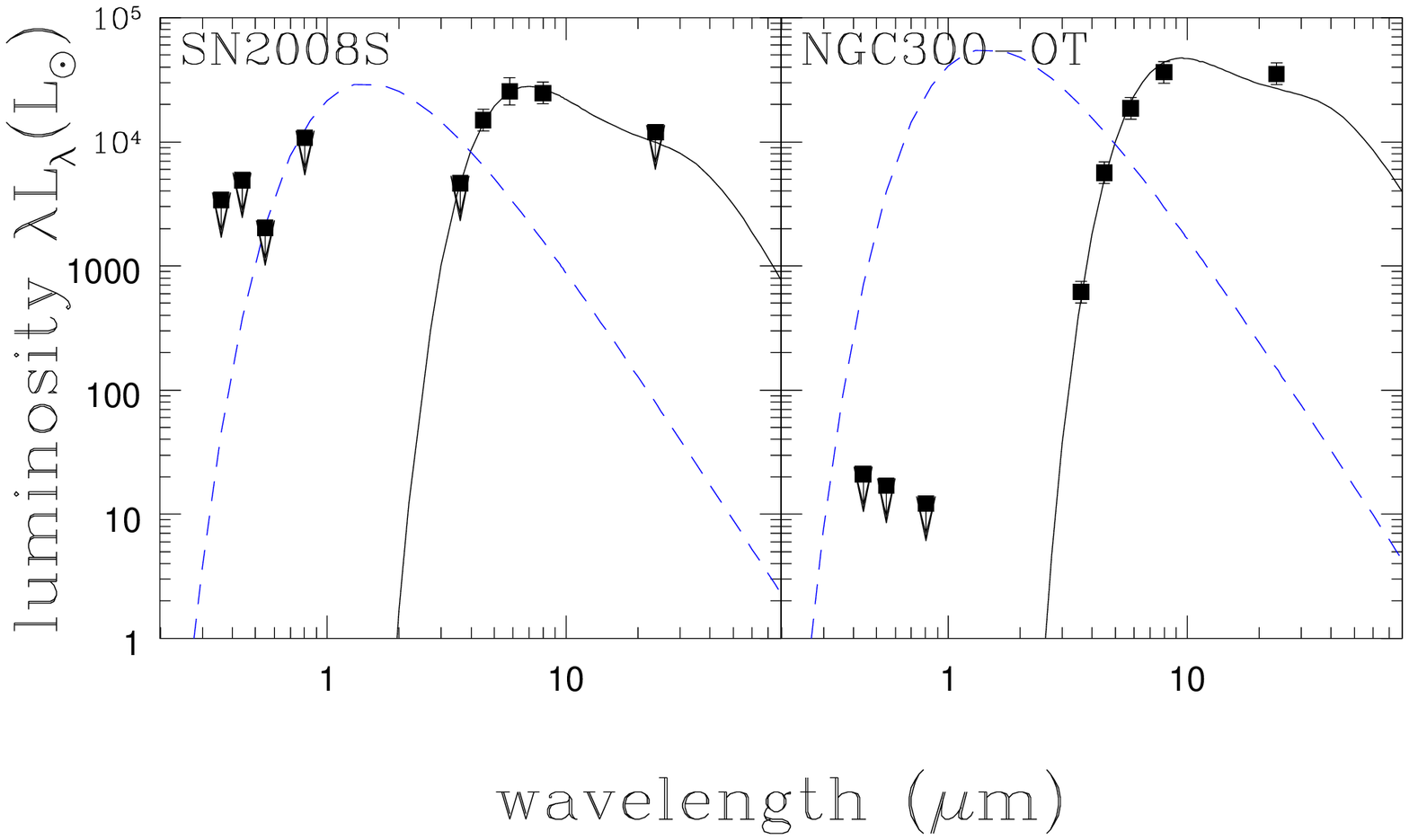}}
\caption{ The spectral energy distributions of the SN~2008S (left) and NGC~300-OT (right)
  progenitors modeled as DUSTY winds.  
  The underlying star is assumed to be a cold, $T_*=2500$~K, extreme AGB
  star, (badly) modeled with the black body SED shown by the dashed lines. Note, however,
   that no particular stellar temperature is required by the models because of the high
  optical depths.  The solid
  curves show the DUSTY wind model for the observed SED assuming an inner edge
  dust temperature of $1500$~K.  The models have visual optical depths of $\tau_V=300$
  and $750$, respectively, and stellar luminosities of $L_*=10^{4.6}L_\odot$ 
  and $10^{4.9}L_\odot$.  
  The SN~2008S data are from \cite{Prieto2008}, the NGC~300-OT mid-IR data are
  from \cite{Prieto2008a} (also \cite{Thompson2009}) and the NGC~300-OT optical limits are from
  \cite{Berger2008} (also \cite{Berger2009} and \cite{Bond2009}).
  }
\label{fig:progenitor}
\end{figure}

\begin{figure}[t]
\centerline{\includegraphics[width=5.5in]{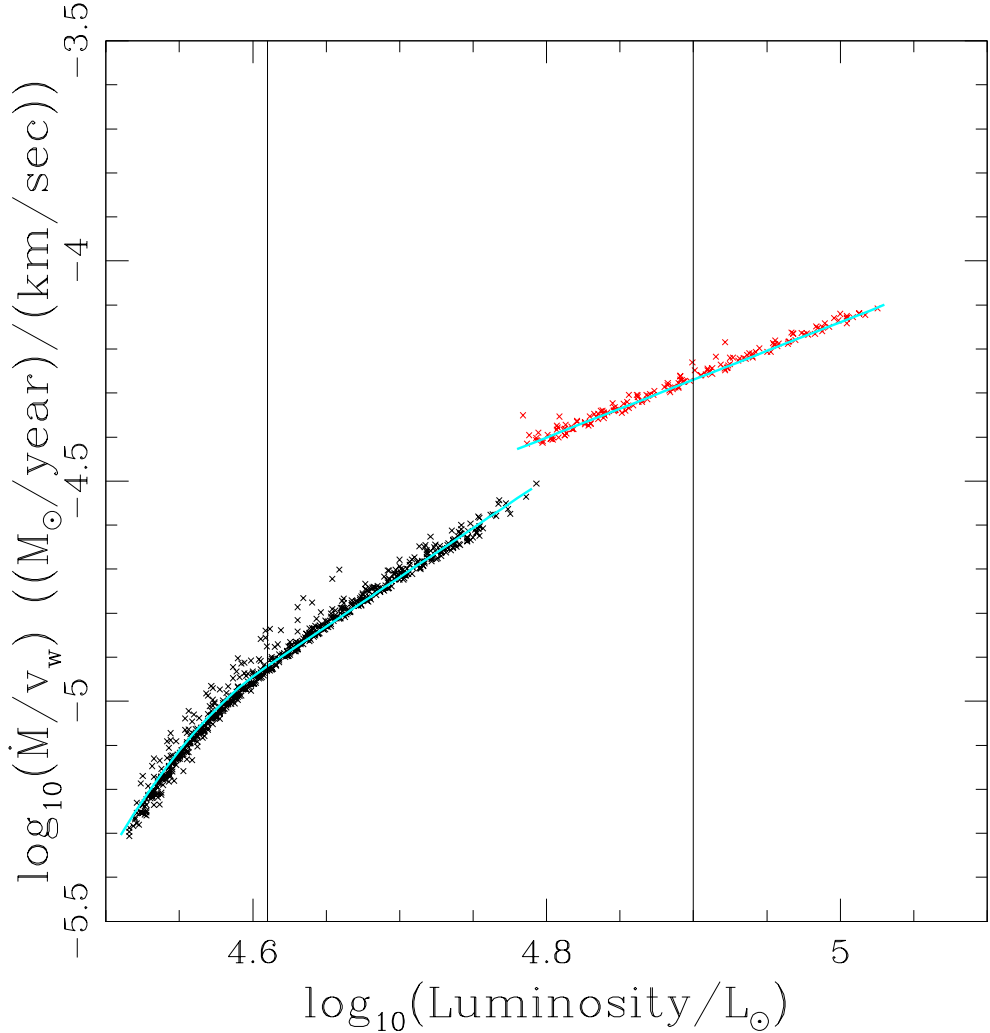}}
\caption{ The stellar luminosity and wind density parameters for DUSTY models
  of the progenitors of SN~2008S (left, black) and the NGC~300-OT (right, red).   
  These include all models roughly consistent with the photometry and having
  inner edge dust temperatures $\geq 1000$~K.  The cyan curve shows the
  wind density predicted from the radius $R(\tau_V=1)$ where the visual
  optical depth is unity and a dust optical depth of $\tau_V=137$~cm$^2$/g. This
  is for DUSTY's default gas-to-dust ratio of $r_{gd}=200$, and the wind density can be
  raised as $\dot{M}/v_w \propto r_{gd}$ by reducing the dust opacity
  per unit mass, $\kappa_V \propto r_{gd}^{-1}$, or {\it vice versa}. 
  }
\label{fig:mwind}
\end{figure}

\begin{figure}[t]
\centerline{\includegraphics[width=5.5in]{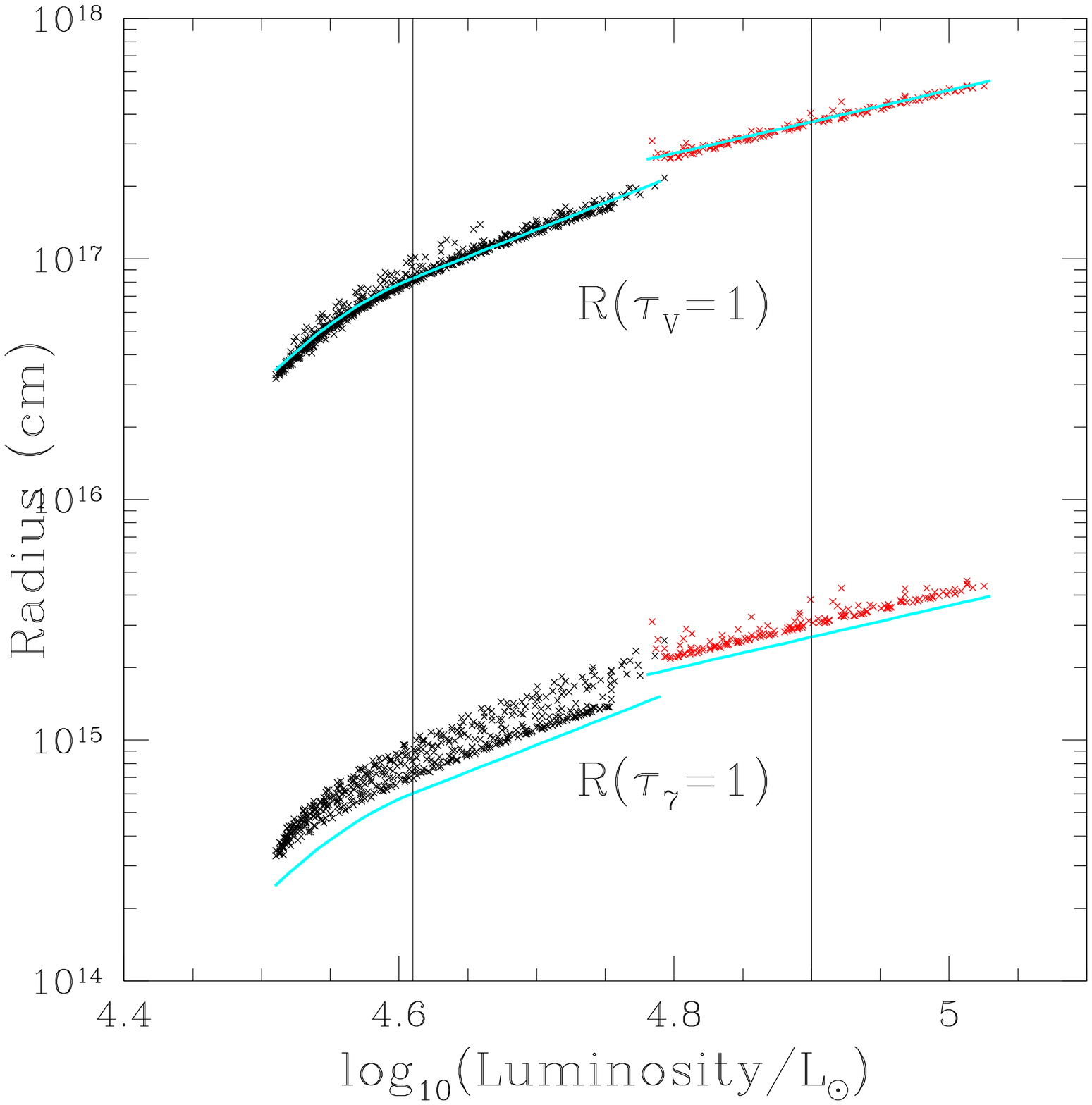}}
\caption{ The radii $R(\tau_V=1)$ and $R(\tau_7=1)$ where the visual optical (upper points)
  and $7\mu$m (lower points) optical depth is unity for the 
  SN~2008S (left, black) and the NGC~300-OT (right, red) progenitor winds.
  The vertical lines show the model likelihood-weighted
  mean luminosities.  The cyan line through the $R(\tau_V=1)$ points shows the
  simple fits given in Eqn.~\ref{eqn:fit1}, while the line just below the 
  $R(\tau_7=1)$ points show the prediction for that radius given a simple
  $\rho \propto 1/r^2$ wind model. 
  }
\label{fig:rwind}
\end{figure}

\section{The Progenitors}
\label{sec:progenitors}

An EAGB star is surrounded by a radiatively accelerated wind of mass loss rate $\dot{M}$ and 
asymptotic velocity $v_w$ (e.g. \citealt{Poelarends2008}, \citealt{Ivezic2010}).  We are only 
interested in the radial regime where the wind has
reached its asymptotic velocity and the density distribution is $\rho_{gas} = \dot{M}/4\pi v_w r^2$.
Thus, one fundamental parameter is the wind density parameter, $\dot{M}/v_w$, which is 
proportional to both the wind optical depth at any wavelength and the luminosity produced by an 
outgoing shock wave (see \S\ref{sec:xray}).  We fit the observed spectral energy distributions (SED) of the
SN~2008S and NGC~300-OT progenitors using the dusty wind models of DUSTY (\citealt{Ivezic1997}, \citealt{Ivezic1999}).  Fig.~\ref{fig:progenitor}
shows the two progenitor SEDs fit as very cool stars ($T_*=2500$~K) surrounded by dense
winds with inner edge dust temperatures of $1500$~K.  

The models are not unique in terms of stellar temperatures or inner edge dust temperatures,
but the stellar luminosity $L_*$ and the density parameter of the wind $\dot{M}/v_w$ are
well-defined, as shown in Fig.~\ref{fig:mwind}.  The SN~2008S progenitor had 
luminosity $L_* = 10^{4.61\pm0.04} L_\odot$ and requires a wind
with $\log(\dot{M}/v_w)= -4.93\pm 0.12$, where $\log \dot{M} \simeq -3.8 \pm 0.1$ 
is in $M_\odot$/year and $v_w \simeq 15 \pm 4$ is in km/s.  The mass loss and
velocity estimates are based on DUSTY's self-consistent wind acceleration models.
The NGC~300-OT progenitor had    
luminosity $L_* = 10^{4.90\pm0.04} L_\odot$ and requires a wind
with $\log(\dot{M}/v_w)= -4.26\pm 0.06$, where $\log \dot{M} \simeq -3.2 \pm 0.1$ 
is in $M_\odot$/year and $v_w \simeq 12 \pm 3$ is in km/s.  The time required
for the wind to expand to these radii is consistent with the $\ltorder 10^4$~year
limit on the lifetimes of the progenitors in this state (\citealt{Thompson2009}).  

As emphasized by \cite{Ivezic2010}, there is a limit to the mass loss rate 
that can be driven by radiation pressure on dust, which they refer to as
the ``reddening bound.''  If these are extreme AGB stars with masses of order $M_*\simeq 9M_\odot$,
the progenitors are near but below this limit for graphitic dust, but above the limit for
silicate dusts.  This adds further support for graphitic over silicate dusts in these 
systems (see \citealt{Prieto2009}, \citealt{Wesson2010}).  The winds of these systems are 
still extreme, and not those of typical AGB stars with dusty winds 
(e.g. \citealt{Matsuura2009}).  But we already know that these are very atypical stars, with
short lifetimes and only a few such systems per galaxy 
(\citealt{Thompson2009}, \citealt{Khan2010}).

The parameter which best scales our later results is the radius at which the visual
optical depth of the progenitor wind is unity, $R(\tau_V=1$).  The DUSTY model estimates of this 
radius are shown in Fig.~\ref{fig:rwind}, and they can be well-approximated by
\begin{eqnarray}
    \log R(\tau_V=1) &= &16.92 + 2.24(\log L_*-4.61) - a (\log L_*-4.61)^2 \\
    \log R(\tau_V=1) &= &17.57 + 1.32(\log L_*-4.90) \nonumber
    \label{eqn:fit1}
\end{eqnarray}
for SN~2008S and the NGC~300-OT, respectively. Here,  $a=16$ if $\log L_* < 4.61$ and $a=0$ 
if $\log L_* >4.61$, and the luminosity ranges are $4.51 < \log L_* < 4.79$ (SN~2008S)
and $4.78 < \log L_* < 5.03$ (NGC~300-OT).
This radius is directly related to the wind density parameter 
\begin{equation}
           { \dot{M} \over v_w } = { 4 \pi R(\tau_V=1) \over \kappa_V}
\end{equation}
with a visual opacity $\kappa_V = 137$~cm$^2$/g, as shown in Fig.~\ref{fig:mwind}. 
The wind density depends on both the bulk density of the dust, where we have
used  $\rho_{bulk} = 2.2$~g/cm$^3$ for graphitic dust (\citealt{Ivezic2010}), and the
gas-to-dust ratio, where we are presenting results for DUSTY's default ratio of
$r_{gd}=200$.  The density parameter, $\dot{M}/v_w$, mass loss rate $\dot{M}$ and wind 
velocity $v_w$ scale as $r_{gd}\rho_{bulk}$, $(r_{gd}\rho_{bulk})^{3/4}$ and 
$(r_{gd}\rho_{bulk})^{-1/4}$, respectively.  More importantly, the models of the progenitor's
SED directly
constrain only $R(\tau_V=1) \propto \dot{M} \kappa_V/v_w$, so we can increase the
gas density of the wind $\dot{M}/v_w \propto r_{gd}$ by reducing the dust opacity
per unit mass $\kappa_V \propto r_{gd}^{-1}$, while leaving the dust radiative
transfer unchanged and ignoring the problem of accelerating the wind.  The default DUSTY value of
$r_{gd}=200$ is larger than the ratio typical of the ISM ($r_{gd} \simeq 100$),
and this is generally true of AGB star winds (\citealt{Ivezic2010}).  While we
leave the gas-to-dust ratio fixed to this default, it does provide a means of
changing the gas density and thus the shock luminosity and X-ray 
emission we discuss in \S\ref{sec:xray} without altering the models for the
SEDs.

Most models of these systems have tried to normalize the parameters using a 
dust photosphere at radius $R_{d*}$ with temperature $T_{d*}$, where 
$L_* = 4 \pi R_{d*}^2 \sigma T_{d*}^4$ (e.g. \citealt{Prieto2008}, \citealt{Thompson2009},
\citealt{Botticella2009}, \citealt{Prieto2009}, \citealt{Berger2009}, while \cite{Wesson2010} fit 
dusty radiation transfer models).   There is no well-defined mid-IR photosphere, however,
because the optical depth varies rapidly with wavelength.  In the DUSTY models, the ratios between the visual
opacity and those at $5$, $7$ and $10\mu$m are $\kappa_{V}/\kappa_5\simeq 79$, 
$\kappa_{V}/\kappa_7 \simeq 138$ and $\kappa_{V}/\kappa_{10} \simeq 200$, respectively.  
If we simply estimate where the radius $R(\tau_7=1)$ at which the $7\mu$m (chosen to lie close to the
peaks of the two SEDs) optical depth is unity, we can illustrate the simple scaling
between radius and wavelength-dependent optical depths.  The simple $\rho\propto 1/r^2$  
model prediction that $R(\tau_7) = R(\tau_V) (\tau_V/\tau_7)(\kappa_7/\kappa_{V})=10^{14.78}$ and $10^{15.42}$~cm
for $\tau_7=\tau_V=1$ and $\kappa_V/\kappa_7=138.5$ agrees well with the DUSTY models, as illustrated in
Fig.~\ref{fig:rwind}. 

\begin{figure}[p]
\centerline{\includegraphics[width=5.5in]{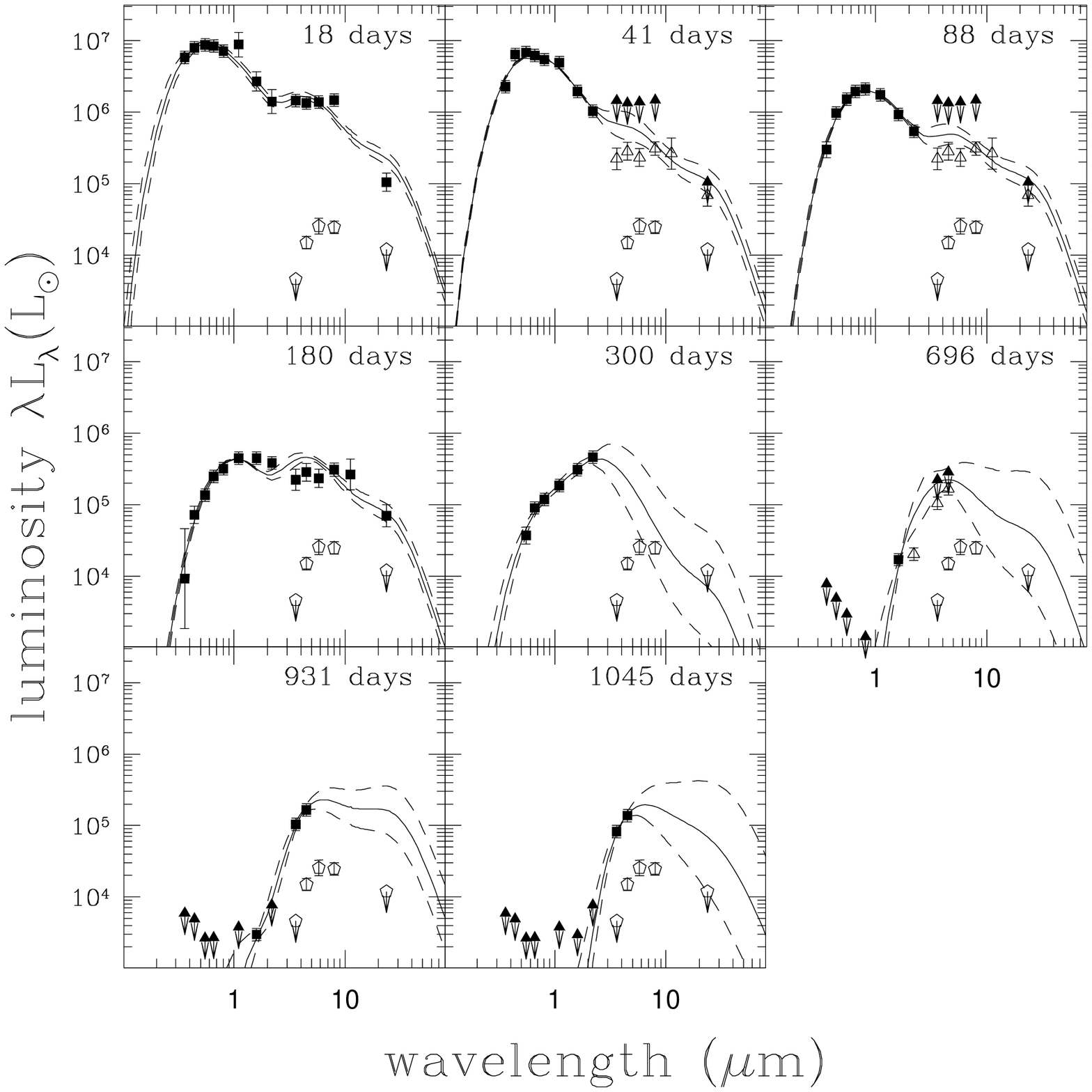}}
\caption{ The evolving SEDs of SN~2008S.  Filled squares
  show measurements at that epoch.  Filled triangles show upper
  bounds either from that epoch or a plausible bound set by the
  other epochs.  Open triangles show plausible lower bounds set
  by the other epochs.  The open pentagons show the SED of the
  progenitor star excluding the optical upper limits for clarity.
  The solid line shows the probability-weighted
  mean SED of the DUSTY models and the dashed line shows the
  dispersion of these SEDs.  Epochs are relative to the same
  reference date as used by \cite{Botticella2009}. 
  }
\label{fig:transient08}
\end{figure}

\begin{figure}[p]
\centerline{\includegraphics[width=5.5in]{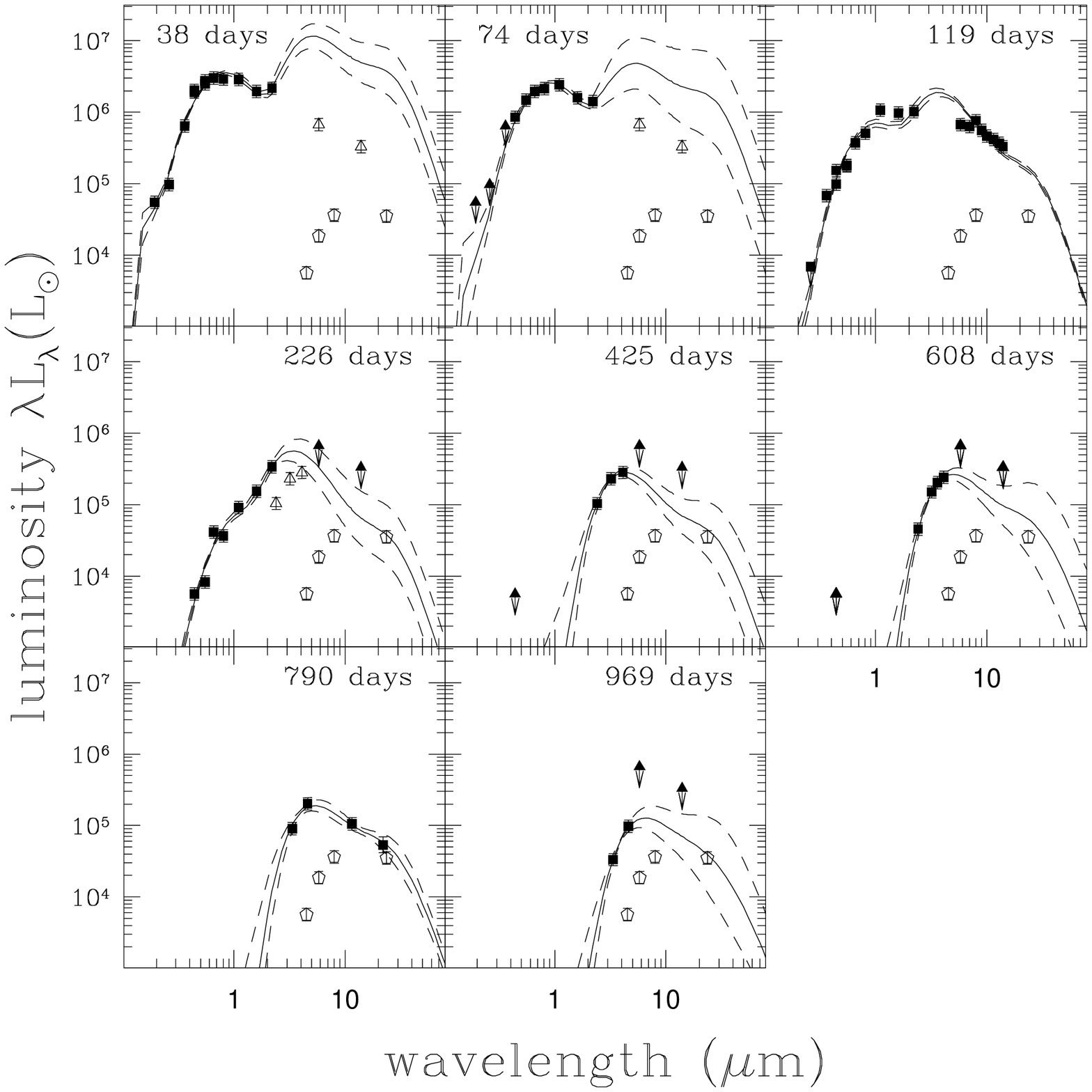}}
\caption{ The evolving SEDs of the NGC~300-OT.  Filled squares
  show measurements at that epoch.  Filled triangles show upper
  bounds either from that epoch or a plausible bound set by the
  other epochs.  Open triangles show plausible lower bounds set
  by the other epochs.  The open pentagons show the SED of the
  progenitor star excluding the optical upper limits for clarity. 
   The solid line shows the probability-weighted
  mean SED of the DUSTY models and the dashed line shows the 
  dispersion of these SEDs.  The optical limit from \cite{Prieto2010a}
  is applied to the last two epochs but lies below the lower edge.
  Epochs are relative to 17 April 2008.
  }
\label{fig:transient300}
\end{figure}

\begin{figure}[p]
\centerline{\includegraphics[width=5.5in]{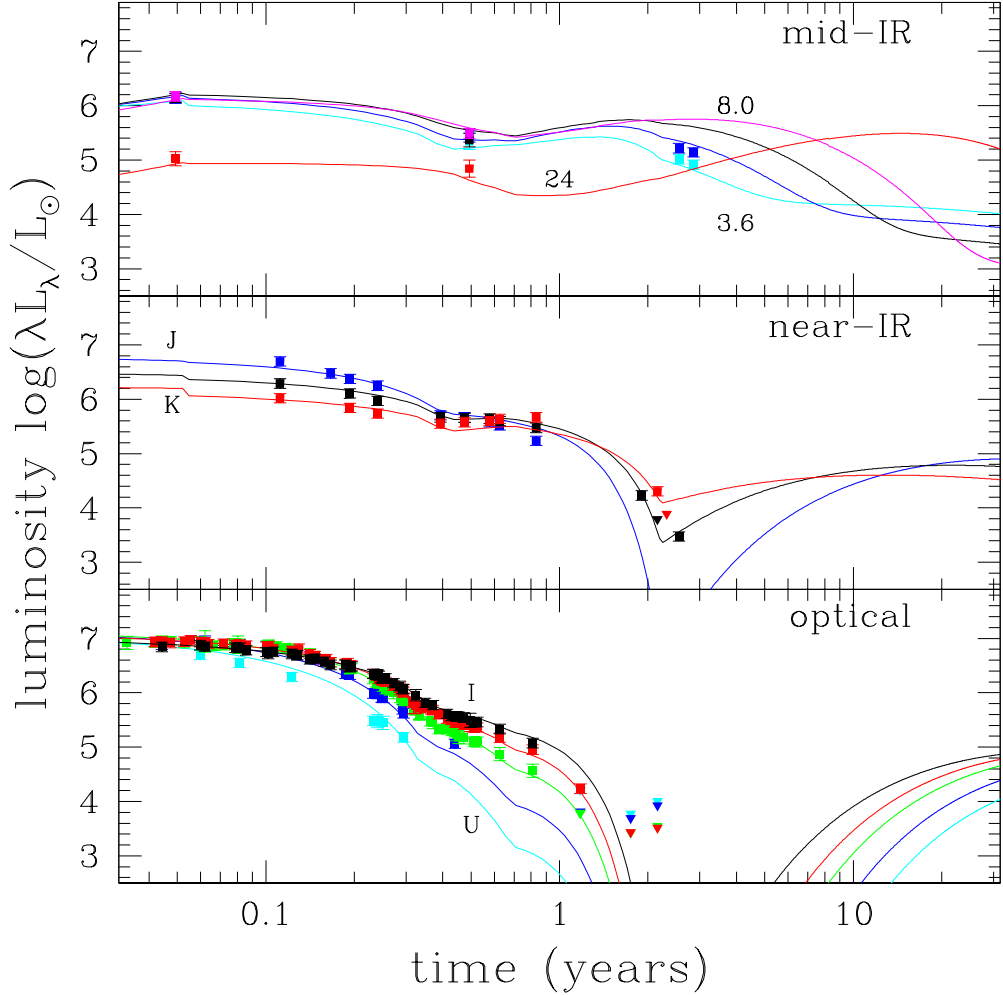}}
\caption{ Simple light curve models for SN~2008S.  The top, middle and bottom panels
  show the mid-IR (IRAC $3.6$, $4.5$, $5.8$, $8.0$ and MIPS $24\mu$m), near-IR
  (J, H and K), and optical (UBVRI) light curves along with the simple model 
  fits.  Measurements are shown by filled squares, upper bounds by triangles.
  The late time optical/near-IR light curves primarily illustrate
  the time scales on which the absorption exterior to the shock becomes
  negligible -- while correct for overall energetics, they are wrong in
  detail because they (incorrectly) assume that the SED of the shock emission 
  is the same as the black body SED of the optical transient.
  }
\label{fig:lc1}
\end{figure}

\begin{figure}[p]
\centerline{\includegraphics[width=5.5in]{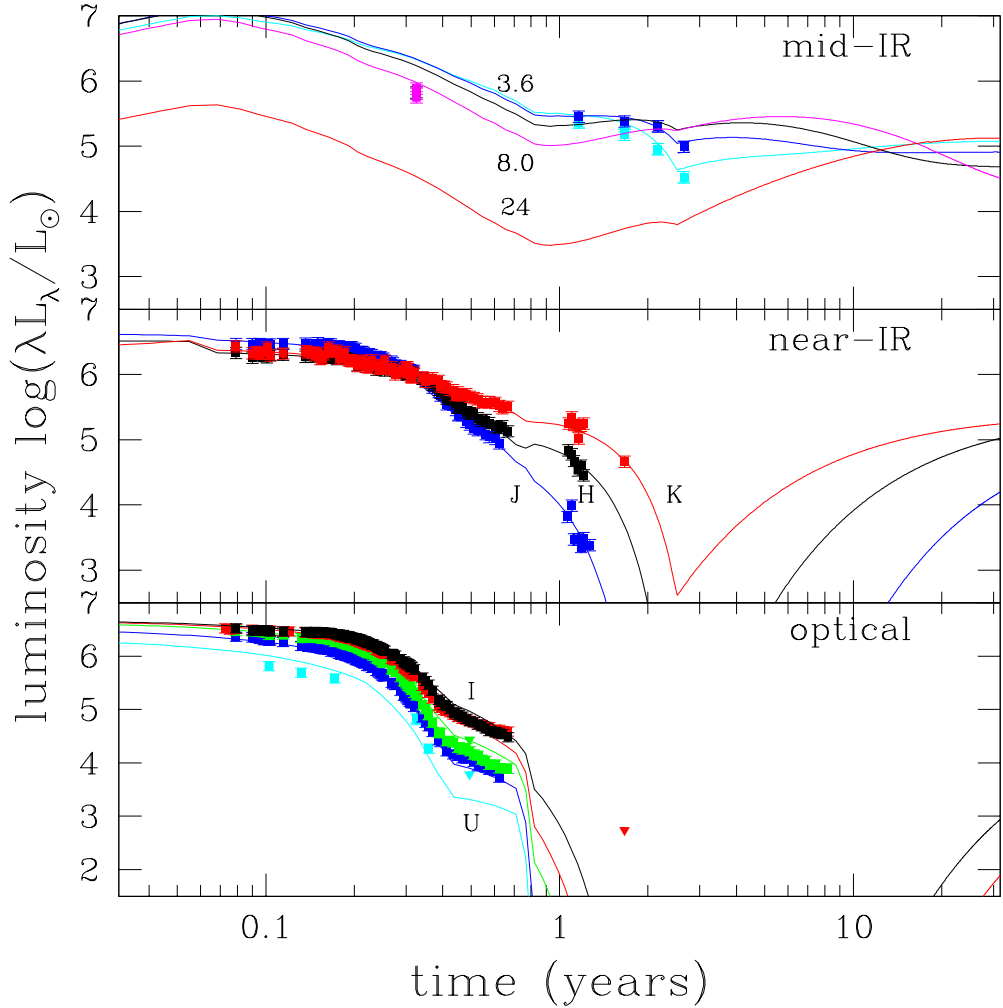}}
\caption{ Simple light curve models for the NGC~300-OT.
   The top, middle and bottom panels
  show the mid-IR (IRAC $3.6$, $4.5$, $5.8$, $8.0$ and MIPS $24\mu$m), near-IR
  (J, H and K), and optical (UBVRI) light curves along with the simple model
  fits.  Measurements are shown by filled squares, upper bounds by triangles. 
  Models allowing modest changes in the transient temperature better
  fit the ``kink'' in the optical light curves.  The same caveats apply
  here as in Fig.~\ref{fig:lc1}. The poorer fit for the R band is probably
  due to H$\alpha$ emission.
  }
\label{fig:lc2}
\end{figure}

\begin{figure}[t]
\centerline{\includegraphics[width=3.75in]{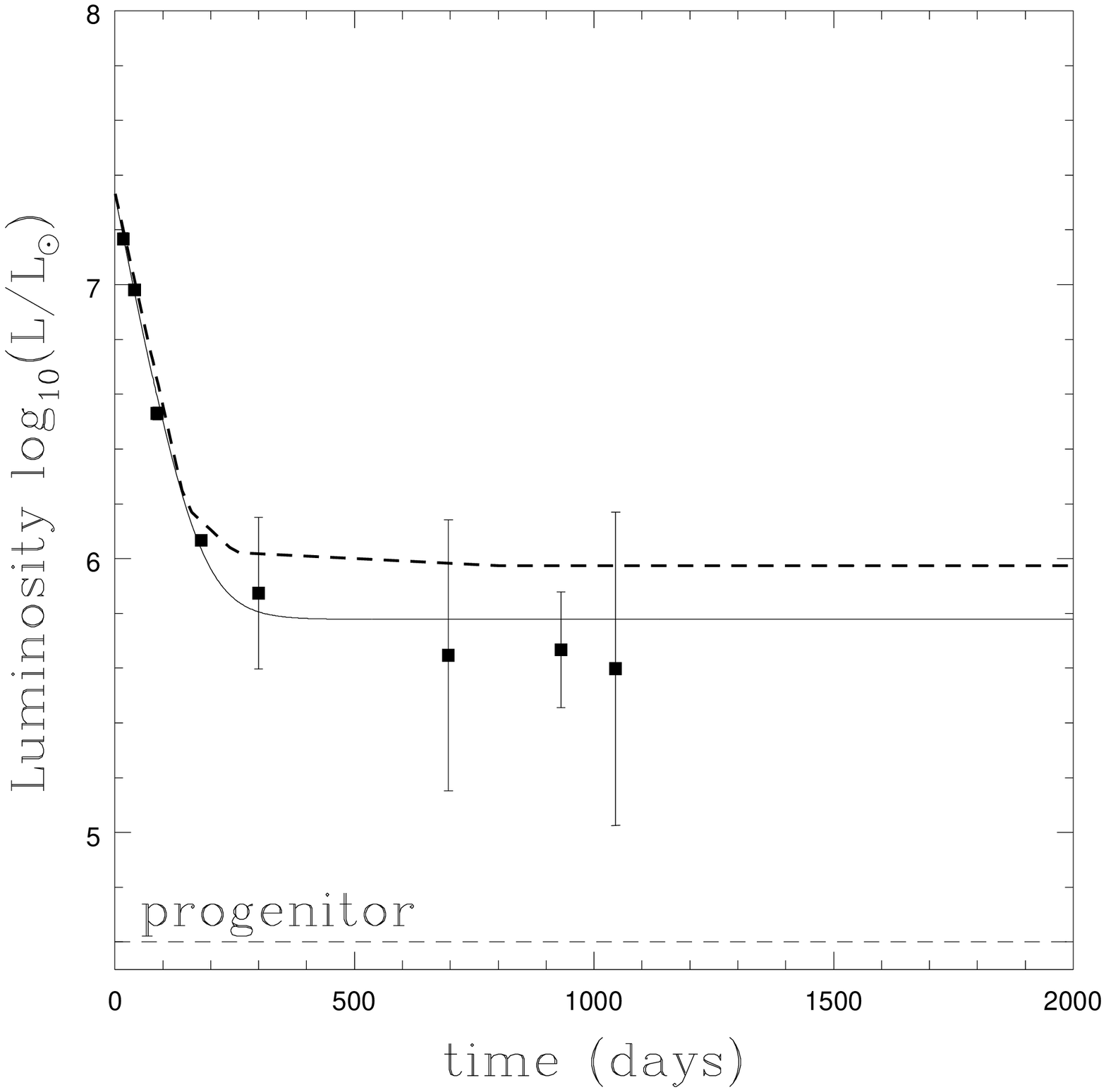}
            \includegraphics[width=3.75in]{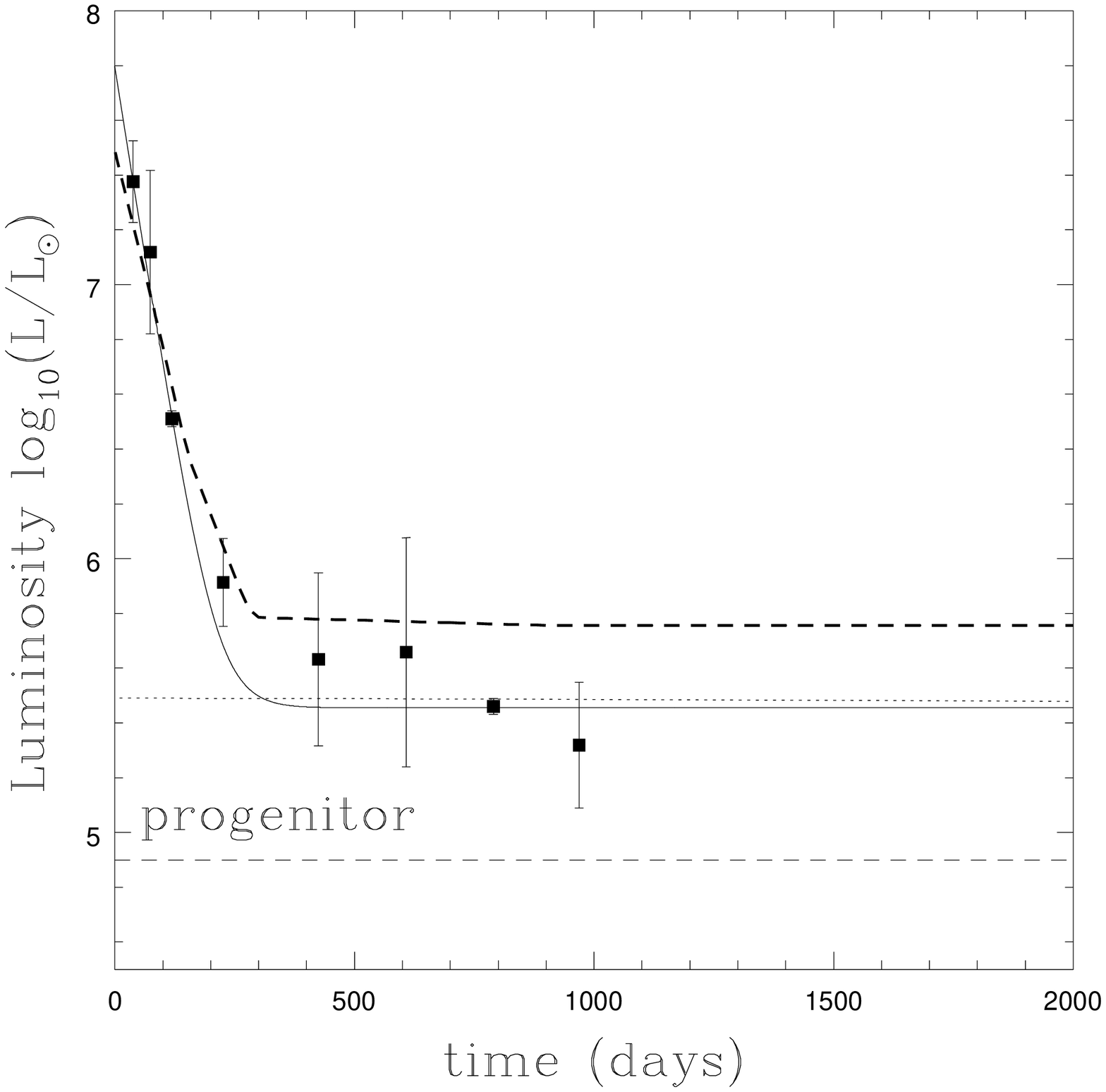}}
\caption{ The luminosity evolution of SN~2008S (left) and the NGC~300-OT (right). The
  points show the luminosity estimates from the DUSTY models.  The solid curve
  shows a simple exponential plus constant model for the luminosity, while the
  heavy dashed curve shows the model from the light curve fits.  The
  dashed curve shows the luminosity of the progenitor.
  }
\label{fig:levolve}
\end{figure}

\begin{figure}[t]
\centerline{\includegraphics[width=3.75in]{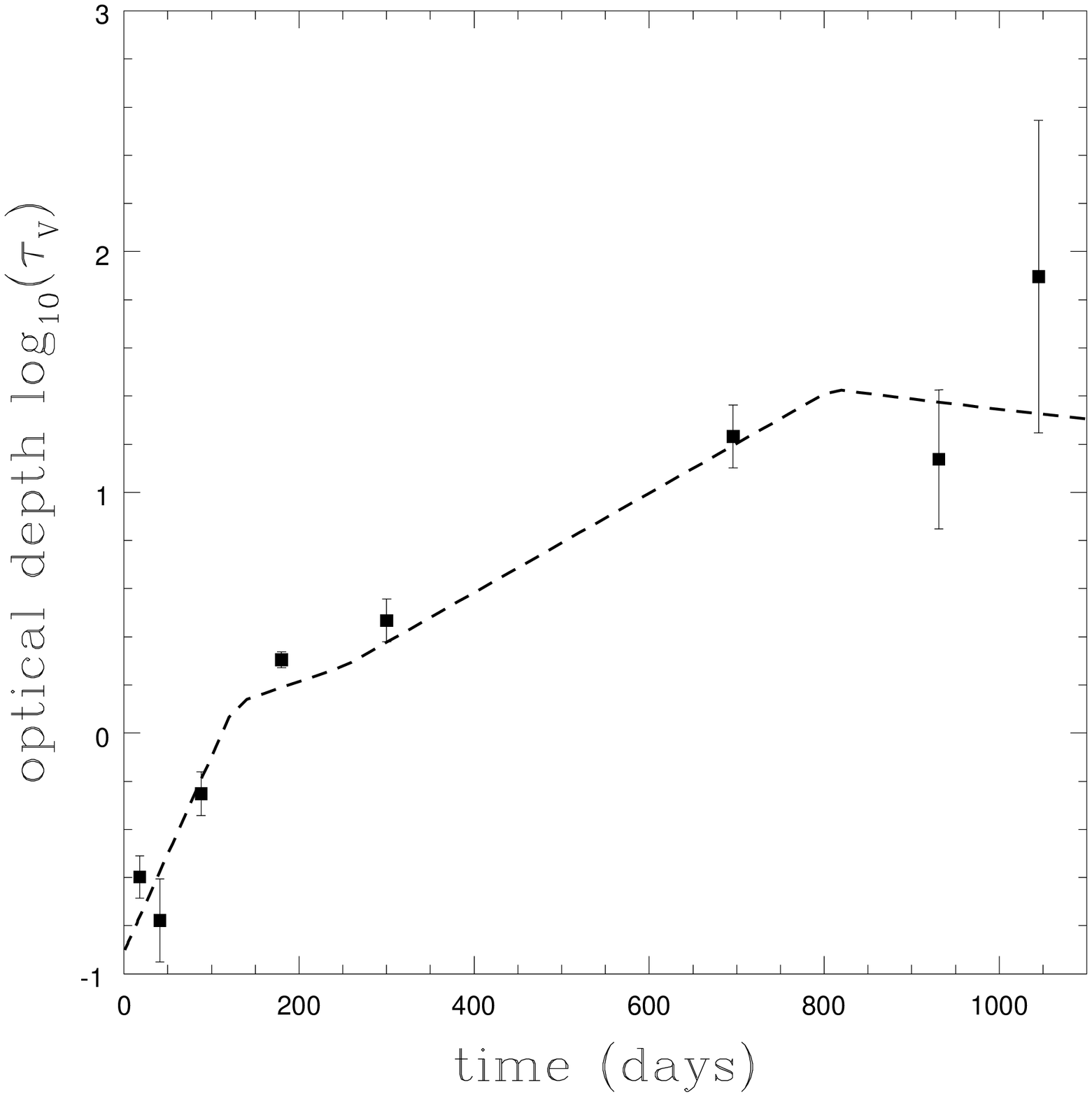}
            \includegraphics[width=3.75in]{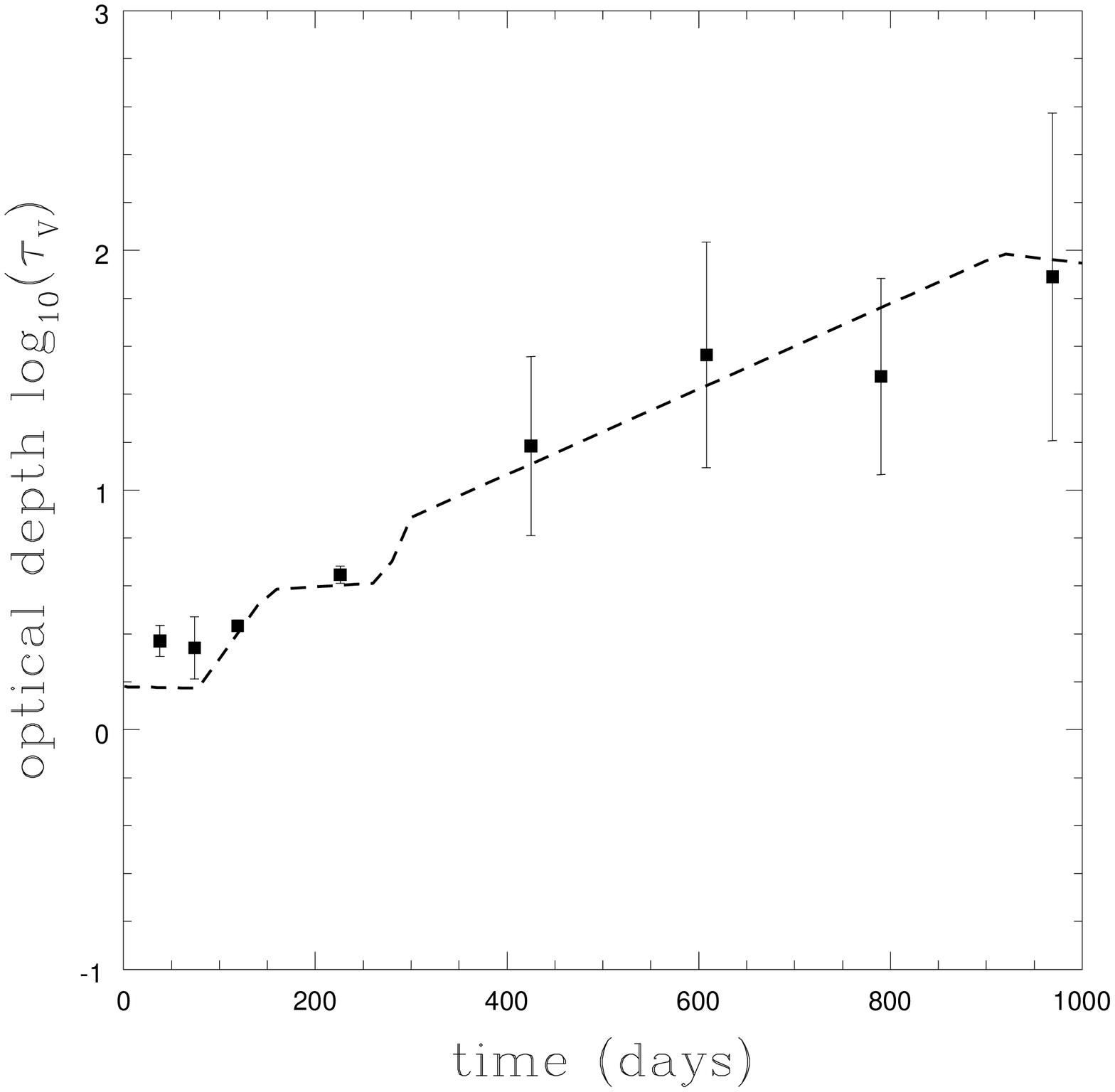}}
\caption{ The optical depth evolution of SN~2008S (left) and the NGC~300-OT (right). 
  The points show the $\tau_V$ estimates from the DUSTY models, while the heavy
  dashed curve shows the results from the light curve fits.  
  }
\label{fig:tevolve}
\end{figure}

\begin{figure}[t]
\centerline{\includegraphics[width=3.75in]{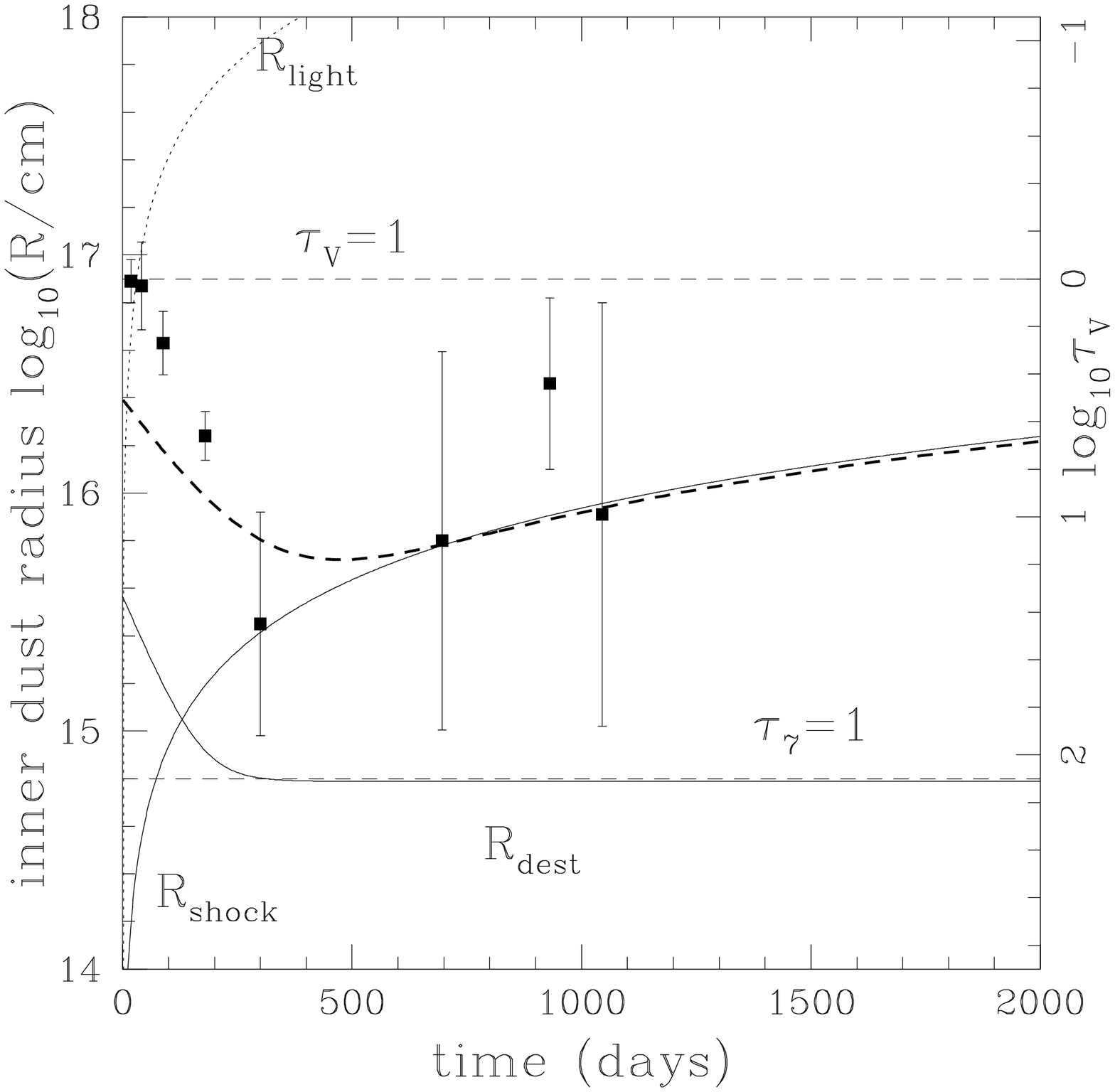}
            \includegraphics[width=3.75in]{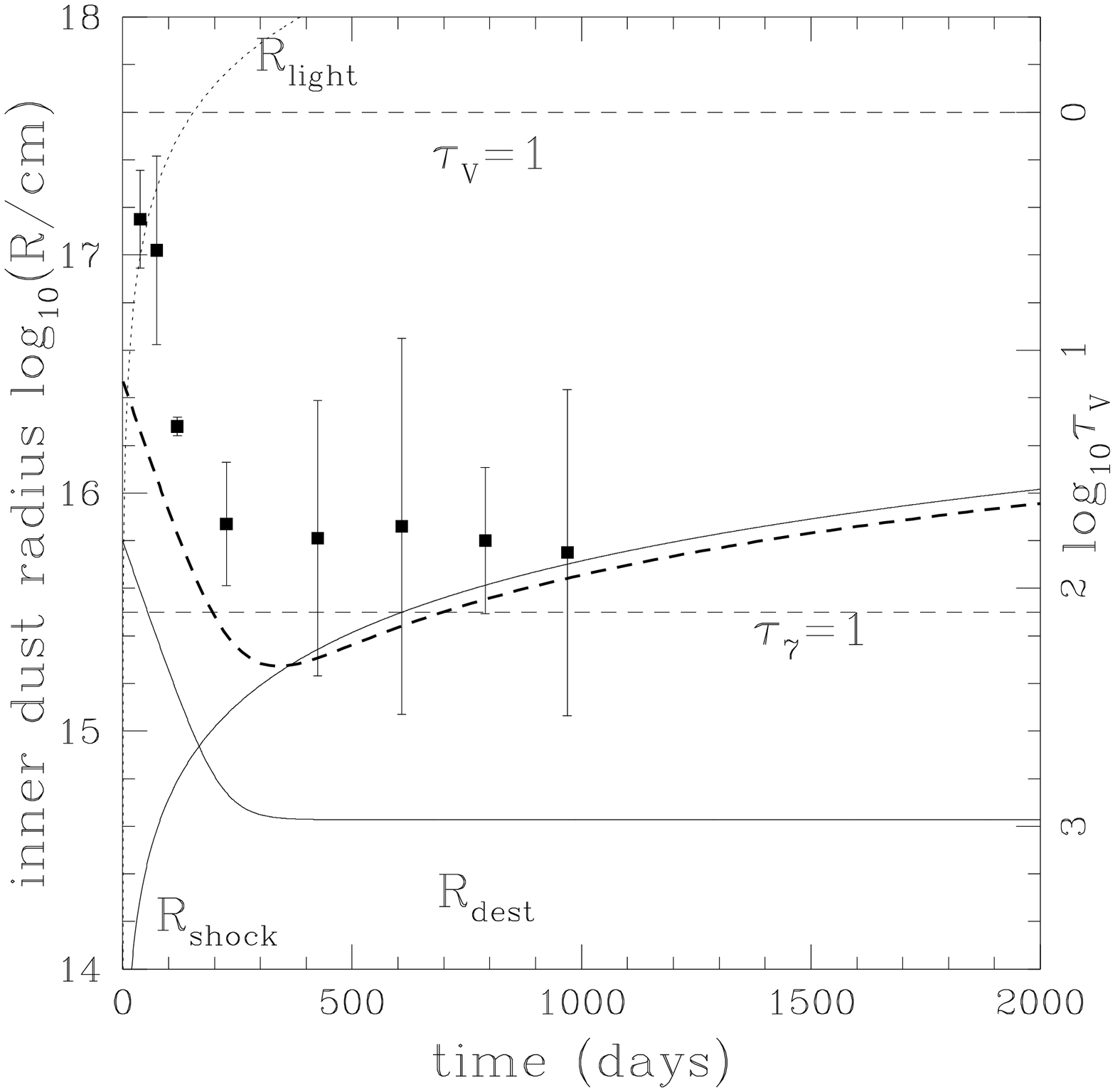}}
\caption{ The dust radius evolution of SN~2008S (left) and the NGC~300-OT (right). The
  points show the radius of the inner edge of the dust from the DUSTY models while 
  the heavy dashed line shows the model from the light curve fits. 
  The right axis shows the visual optical depth scale of the progenitor wind,
  where the horizontal dashed lines mark the radii where the visual ($\tau_V=1$)
  and $7\mu$m ($\tau_7=1$) optical depths to infinity were unity.
  The solid lines show the small grain dust destruction radius $R_{dest}$
  from Eqn.~\ref{eqn:dform} predicted from the luminosity model in Fig.~\ref{fig:levolve}, and the 
  shock radius $R_{shock}=v_s t$ assuming velocities of $v_s=1000$ and
  $600$~km/s for SN~2008S and the NGC~300-OT, respectively.  The dotted
  line shows the light radius $R_{light}=c t$.  Epochs above this line will
  be affected by finite light crossing times, although not significantly on
  the logarithmic scales of Figs.~\ref{fig:levolve}, \ref{fig:tevolve} and
  \ref{fig:revolve}.
  }
\label{fig:revolve}
\end{figure}

\section{The Evolution of the Transients}
\label{sec:transients}

The next step is to characterize the evolution of the transients.  Figs.~\ref{fig:transient08} and 
\ref{fig:transient300} show snapshots of the evolution of the SEDs of SN~2008S and
the NGC~300-OT, and Figs.~\ref{fig:lc1} and \ref{fig:lc2} show the evolution of the optical, near-IR and
mid-IR light curves.  The data for SN~2008S comes from \cite{Botticella2009},
\cite{Smith2009}, \cite{Wesson2010}, \cite{Prieto2011} and \cite{Hoffman2011},  while that for the NGC~300-OT 
comes from \cite{Bond2009}, \cite{Berger2009}, \cite{Prieto2009}, \cite{Prieto2010a}, \cite{Ohsawa2010}
and  \cite{Hoffman2011}.  We will
model these SEDs with DUSTY, ignoring light travel times (``dust echoes'', \citealt{Wright1980}, \citealt{Dwek1983}), 
as these have already been
considered in some detail by \cite{Wesson2010}, and we are primarily interested in the later time behavior
when they are unimportant.  This does mean that our very early time models will somewhat mis-characterize
the dust properties, where the general sense is to underestimate the radius of the dust and its optical depth.
However, on the logarithmic scales in which we are interested, these are minor corrections, and
we will incorporate echoes into our light curve models to explore their effects.  
We can see from Figs.~\ref{fig:transient08} and \ref{fig:transient300}
that the transients, like the progenitors, cannot be well-characterized without mid-IR
observations.  In fact, in the most plausible models, the NGC~300-OT peaked in the mid-IR 
rather than the optical.  In both cases, as the luminosity drops, the SED shifts from the 
optical to the mid-IR and the transients again become fully obscured (\citealt{Prieto2010a},
\citealt{Prieto2011}). 

The relevant variables for describing the transients are the (bolometric) luminosity and temperature 
of the transient and the characteristic radius and optical depth of any surrounding dust.  We first
estimated these quantities using DUSTY models of the individual epochs shown in Figs.~\ref{fig:transient08}
and \ref{fig:transient300}.  DUSTY's structure makes it difficult to produce an evolving series
of models to fit the light curves, so we fit the light curves using simpler physical models 
constrained by both the DUSTY models and the light curves.  In these models for the evolution of
the light curves, the transients
are modeled with a varying luminosity, $L(t)$ and temperature $T(t)$ black body transient, obscured
by a time varying dust optical depth $\tau_V(t)$ located at radius $r_d(t)$.  The shock velocities 
were constrained by the estimates of $v_s=1100$ and $560$~km/s for SN~2008S and the NGC~300-OT
from \cite{Smith2011}.
The luminosity evolution can almost be modeled as an exponential decline plus a 
constant, $L = L_0 \exp(-t/t_0) + L_1$, and the temperature $T(t)$ can almost be held constant,
but this proved to be inadequate for detailed
models of the light curves.  Thus, we modeled the logarithms of the luminosity, temperature and 
optical depths as piecewise linear functions.  We also included an initial unobserved luminosity
spike that heats the dust at radius $r_d(t=0)$ to a destruction temperature of $T_{dest}=1500$~K.
The dust emission is then time averaged to include the effects of ``dust echoes''
in the light curve models.  The full details of the model are presented in the
Appendix.  Obviously, this model has some {\it ad hoc} features, but it 
produces remarkably good fits to the light curves given its simplicity,  
 as shown in Figs.~\ref{fig:lc1} and \ref{fig:lc2}.  
The modest mismatches of the DUSTY and semi-analytic estimates of the dust radii
and optical depths illustrate some of the biases created by the approximations.

Fig.~\ref{fig:levolve} summarizes the luminosity evolution of the transients.  The luminosities
drop exponentially in the early phases and then settle at a nearly constant plateau, given the 
uncertainties in reconstructing the luminosity with variable wavelength coverage. 
If we approximate the luminosity as $ L = L_0 \exp(-t/t_0) + L_1$, we find
$L_0 \simeq 10^{7.3} L_\odot$, $L_1 \simeq 10^{5.8}L_\odot$ and $t_0\simeq 48$~days for SN~2008S
and $L_0 \simeq 10^{7.8} L_\odot$, $L_1 \simeq 10^{5.5}L_\odot$ and 
$t_0\simeq 39$~days for the NGC~300-OT.  The uncertainties are almost entirely systematic, so we do not
report formal statistical uncertainties.  Thus, the NGC~300-OT transient was actually
brighter, rather than fainter than SN~2008S, and radiated more energy in the transient
phase ($E\simeq 3 \times 10^{47}$~ergs for SN~2008S versus $E \simeq 8 \times 10^{47}$~ergs
for the NGC~300-OT). Here we have calculated only the contribution from the exponential
term.  The energy associated with the constant term is unimportant on time scales of $t_0$,
but matches it if the emission continues for 4 and 21~years for SN~2008S and the NGC~300-OT,
respectively.  SN~2008S has nearly reached this point, while the time scale is much longer
for the NGC~300-OT because of its stronger transient and lower plateau luminosity. 
The inferred peak luminosities, $L_0 \simeq 8 \times 10^{40}$
and $2 \times 10^{41}$~ergs/sec are comparable to those of low luminosity
Type~IIP SNe (e.g. \citealt{Pastorello2004}). In the plateau phase, SN~2008S is brighter than 
the NGC~300-OT, and both transients remain significantly brighter than their 
progenitor stars.   In both cases, the energetics and luminosities derived from
traditional optical approaches are highly misleading -- for example, the luminosity
history of these transients including the mid-IR emissions is very different from
the histories based only on optical observations presented by \cite{Smith2011}.

Fig.~\ref{fig:tevolve} shows the evolution of the optical depths.  Where
the progenitors had enormous dust optical depths, the transients at peak had little
obscuration, only $\tau_V \simeq 0.3$ for SN~2008S and $\tau_V \simeq 2$ for the 
NGC~300-OT (\citealt{Prieto2008},
\citealt{Botticella2009}, \citealt{Bond2009}, \citealt{Wesson2010}).   Note the congruence between the optical depth estimates
at peak in Fig.~\ref{fig:tevolve} and the optical depth of the progenitor wind
outside the estimate of the initial radius of the dust in Fig.~\ref{fig:revolve}. 
As the transients evolve, the dust optical depth steadily
rises, with $\log \tau_V \simeq 1.9 \pm 0.7$ for the final DUSTY epoch in both cases,
where the large uncertainties are due to the incomplete wavelength coverage in 
the final epochs.  The optical depth does not, however, initially rise to the total optical
depth of the progenitor wind outside the estimated emission radius, but it does appear to 
steadily approach such optical depths at later times.  As the dust
destruction radius recedes, the dust appears to reform from the outside
inwards, but some time is required
before the optical depth returns to the full pre-transient values.  The time scale for
restoring most of the dust opacity outside the expanding shock appears to be approximately $2$-$3$ years in both 
systems.

In the DUSTY models, the preferred transient radiation temperatures drop from 7000-8000~K
in the first month to 5500-6500~K after 100 days, similar to the evolution in the
semi-analytic models.
These tend to be somewhat warmer than previous models where the 
optical temperature was estimated by fitting the emission without a self-consistent
model for the dust absorption.  At later times, the heavy obscuration makes it 
impossible to estimate the intrinsic temperature of the transient.
As the dust reforms and the characteristic radius of the dust
shrinks, the characteristic dust temperature rises with a peak temperature after 
roughly 8-9~months.  This supports
the strong near-IR emission in both sources even as the optical emission is collapsing.
Fig.~\ref{fig:revolve} shows that the initial transient must destroy the dust
out to a radius of order $10^{17}$~cm, and it then seems to reform from the outside
inwards as the dust destruction radius retreats.  Eventually the radius of the dust 
approaches the shock radius and the SED models are consistent with the dust radius
starting to expand outwards again.  

While there is no ambiguity about the presence of dust near the peak of SN~2008S
due to the early-phase mid-IR detections of \cite{Wesson2010}, our claim of a significant 
dust optical depth at the peak of the NGC~300-OT differs strongly from the view
of \cite{Berger2009} that there was no dust optical depth near the peak and perhaps
weakly with the modest obscuration proposed by \cite{Bond2009}.  The first
argument in \cite{Berger2009} is that their optical/UV SED at 43 days is well-fit 
by a $\sim 4700$~K black body with no extinction based on the photometry from
z-band to the SWIFT UVW1 (2500\AA) band.  In Fig.~\ref{fig:transient300} we
have extended this wavelength baseline to include the near-IR data from \cite{Bond2009}
and shifted to a slightly earlier date (25 May instead of 31 May) in order to include 
the shorter wavelength SWIFT UVW2 (1900\AA) measurement from \cite{Berger2009} with less
of a correction based on the evolution of the UVW1 magnitudes.  
Compared to a cool, $4700$~K black body, the SED has both
a near-IR and UVW2 excess, which the DUSTY models resolve by using a hotter 
($\sim 9000$~K) spectrum with an optical depth of $\tau_V \simeq 2.3\pm0.3$.  
This optical depth is also the amount required to reproduce the Balmer 
decrement observed by \cite{Berger2009}.  \cite{Bond2009} interpret the 
spectra as suggesting that the photosphere has the intrinsic color/temperature 
of an F star, which then suggests $\tau_V\simeq 1$, but they did not directly
fit the SEDs.


\cite{Berger2009} argue against any surviving dust
based on the absence of significant Na~I~D absorption in their spectra.  
The lack of absorption is remarkable because the total sodium column
density of the circumstellar medium is of order $10^{15-16} \tau_V$~cm$^{-2}$,
where $\tau_V \gtorder 10^2$ is the visual optical depth of the material obscuring 
the progenitor.  Unsaturated Na~I~D absorption lines mean that all
the sodium exterior to the expanding shock has to be photoionized independent 
of the survival of any dust.   The flux required to photoionize a $\rho \propto 1/r^2$ wind
from $r_{in}$ to $r_{out}$ scales as $Q = Q_0 (1-r_{in}/r_{out})$ where for 
pure hydrogen $Q_0 = \dot{M}^2 \alpha_B / 4 \pi v_w^2 m_p^2 r_{in}$, which
is very different from the $Q \propto r_{out}^3$ scaling of a uniform density
medium (e.g. \citealt{Fransson1982}).  In a wind, once there are enough ionizing photons, $Q\simeq Q_0$, to ionize 
the inner parts of the wind, virtually no additional ionizing photons are 
required to ionize the entire wind.  For example, at 100 days in the NGC~300-OT, 
the shock radius of $ \sim 10^{14.6}$~cm is well inside the progenitor's dust 
``photosphere'' at $ \sim 10^{15.6}$~cm, so the sodium in the dense parts of the
wind has to be photoionized.  Expanding the sodium Str\"omgren sphere from
the dust ``photosphere'' to infinity then only requires $\sim 10$\% more 
ionizing photons.  Moreover, while the transients
cannot keep hydrogen and helium fully ionized due to their low radiation 
temperatures, they are hot enough to produce sodium ionizing photons (see
\S\ref{sec:reform}), thereby allowing hydrogen and helium to recombine
while keeping sodium photoionized.  This leads to very long sodium recombination
times due to the lack of electrons, making it sill easier to keep sodium photoionized. 
Thus, at the phases of the spectroscopic observations, we expect no Na~I~D absorption 
from the wind, and the weak Na~I~D absorption found by \cite{Berger2009} is 
due to a small amount of interstellar absorption.

\section{A Physical Model for the Transients}
\label{sec:consequences}

All these features of the evolution of the transient can be explained as a consequence
of an explosive transient from a red supergiant surrounded by a dense, dusty
wind.  We assume a wind normalized by the progenitor properties
from \S\ref{sec:progenitors} and shock velocities near the values of $v_s=1100$
and $560$~km/s proposed by \cite{Smith2011} for SN~2008S and the NGC~300-OT. 
In \S\ref{sec:destroy} we consider the implications of the dust radius and
optical depths shortly after the peak of the transient, and find that it
requires an explosive transient from a red supergiant.  In \S\ref{sec:reform}
we consider recombination, cooling and dust reformation in the progenitor
wind following this luminosity peak.   In \S\ref{sec:xray}
we explain the roughly constant late time luminosity as emission from an
expanding shock propagating through the dense, progenitor wind.  
Eventually, the
wind column density outside the shock is low enough for the X-rays to 
be visible, but at early times we expect neither detectable X-ray nor radio
emission from the shock.  Finally, in \S\ref{sec:optir} and \S\ref{sec:survivor}
we discuss the evolution of the optical, near-IR and mid-IR emissions and the potential
detectability of surviving progenitor stars.   

The X-ray and UV emission produced by an SN shock wave propagating through
circumstellar material has been considered extensively (e.g.
\citealt{Chevalier1982}, \citealt{Fransson1982}, 
\citealt{Lundqvist1988}, \citealt{Chugai1992}, \citealt{Chevalier1994},
\citealt{Chugai1994}).  Most of these models were developed for normal
supernovae with early and long-lived X-ray emission (e.g. SN~1978K,
\citealt{Schlegel1999} or SN~1980K, \citealt{Canizares1982}).  Our
outline of a model for SN~2008S and the NGC~300-OT are simple versions
of these earlier models.  The resulting expectations only differ
from these models because these transients have significantly 
lower luminosities and shock velocities and are expanding
into exceptionally dense, dusty winds.

\subsection{Dust Destruction, Shock Breakout and the Progenitor Radius}
\label{sec:destroy}

In addition to showing the estimates for the typical radius of the dust emission,
Fig.~\ref{fig:revolve} also shows the radii corresponding to the dust ``photosphere''
of the progenitor at $7\mu$m, $R(\tau_7=1)$, and the radius where the visual
optical depth becomes unity, $R(\tau_V=1)$, as compared to the dust destruction
radius predicted by the simple exponential plus plateau models for the bolometric
luminosity shown in Fig.~\ref{fig:levolve}.  As already noted by \cite{Botticella2009},
\cite{Prieto2009}, \cite{Wesson2010} and \cite{Ohsawa2010}, and illustrated by
our models in Fig.~\ref{fig:revolve}, the surviving dust lies in the un-shocked
wind material and so can only have been destroyed radiatively.  Moreover, the
required dust destruction radii are an order of magnitude larger then the 
destruction radius predicted from the observed transient luminosities, as also shown in
Fig.~\ref{fig:revolve}.  The observed transients could only destroy the dust
out to radii near the mid-IR ``photospheres'' of the progenitors, which would have 
left the transients heavily obscured at peak with $\tau_{peak} > 10$ instead 
of $\tau_{peak} \simeq 1$.

This leads us to the conclusion that the transients must have had short, un-observed, 
high luminosity spikes, and were thus explosive transients.  If $\tau_{peak}$ is
the visual extinction at the peak of the transient, then we must destroy the dust
out to the radius $R_{dest} \simeq \tau_{peak}^{-1} R(\tau_V=1)$ where $R(\tau_V=1)$ is 
the radius in the progenitor wind where the visual optical depth was unity.  The luminosity needed 
to destroy the dust out to radius $R_{dest}$ is
\begin{equation}
   L_{peak} \simeq 16 \pi \sigma T_{dest}^4 R_{dest}^2 Q_{rat} = 
      4 \times 10^{10} \tau_{peak}^{-2} \left( { Q_{rat}^{1/4} T_{dest} \over 1500~\hbox{K} }\right)^4 
    \left( { R(\tau_V=1) \over 10^{17}~\hbox{cm} } \right)^2
         L_\odot,
\end{equation}
where $T_{dest}\simeq 1500$~K is the dust destruction temperature and $Q_{rat}=Q(T_{dest})/Q(T_{peak})$ is the 
ratio of the Planck-averaged absorption efficiencies at the temperatures $T_{peak}$ of the break out radiation
and $T_{dest}$ (e.g. \citealt{Dwek1985}, \citealt{Draine1984}).  $Q_{rat}$ depends on grain size because the
smaller grains reach higher temperatures for a given radiation field, so for a more accurate estimate of the
required luminosities we determined the fraction of the dust destroyed for a \cite{Mathis1977} distribution of 
\cite{Draine1984} graphitic dust with the $0.005 < a < 0.25\mu$m size distribution used by DUSTY. A transient peak 
with luminosity $L_{peak}$ and radiation temperature $T_{peak} = 50000$~K destroys 50\% of the dust opacity to
radius
\begin{equation}
      R_{d50} \simeq 6 \times 10^{16} \left( { L_{peak} \over 10^{10} L_\odot }\right)^{1/2} 
                   \left( { T_{dest} \over 1500 } \right)^{-5/4}~\hbox{cm}
\end{equation}
and 90\% of the dust opacity to $R_{d90} \simeq R_{d50}/2$.  Since we require 
$R_{d50} \simeq R(\tau_V=1)/\tau_{peak} \simeq 10^{17}$~cm, we must have 
\begin{equation}
        L_{peak,50} \simeq 3 \times 10^{10} L_\odot \left( { R(\tau_V=1) \over \tau_{peak} 10^{17}~\hbox{cm} } \right)^2
                \left({ T_{dest} \over 1500~\hbox{K} }\right)^{2.5} L_\odot
   \label{eqn:rd50}
\end{equation}
with the luminosity $L_{peak,90}$ needed to destroy 90\% of the opacity being roughly 4 times higher.
{\it Such a high peak luminosity can only be explained by the luminosity produced from a shock breaking out of 
the stellar photosphere, which in turn requires that these were explosive transients rather than some form of
radiation pressure driven eruption.}

Producing such a high luminosity with a shock breakout also implies that the progenitor stars were cold,
red supergiants.  We show this following the approximate breakout model of \cite{Ofek2010} (following \citealt{Falk1977},
\citealt{Waxman2007}). The shock breaks out when the radiative diffusion time scale matches the 
expansion time scale, $t_{diff} = \kappa_T \rho_s r_s^2/c = t_{exp} = r_s/v_s$, where $\kappa_T \simeq 0.4$~cm$^2$/g is the Thomson 
opacity.  This determines the density $\rho_s$ of the breakout region in terms of the break out radius $r_s$,
\begin{equation} \rho_s = { c \over v_s \kappa_T r_s }.
\end{equation}
The post-shock region is radiation dominated, so the radiative energy density has to match the energy density of the shock,
$ a T_s^4 = (7/2) \rho_s v_s^2$, where we have assumed the density compression for $\Gamma=4/3$.  Thus, 
\begin{equation} 
      \sigma T_s^4 = { 7 \over 8 } { v_s c^2 \over \kappa_T r_s }.
      \label{eqn:tbshock}
\end{equation}
By setting the break-out luminosity $4 \pi \sigma T_s^4 r_s^2 f = L_{peak,50}$, we find that the radius is 
\begin{equation}  
   r_s \simeq 5 \times 10^{13} 
              \left( { R(\tau_V=1) \over \tau_{peak} 10^{17} \hbox{cm} } \right)^2 
              \left( { T_{dest} \over 1500~\hbox{K} }\right)^{5/2}
                 \left( { 1000~\hbox{km/s} \over f v_s } \right)~\hbox{cm}.
\end{equation}
This is relatively crude estimate, so we have introduced the dimensionless factor $f$ to monitor the effects of the 
approximation on other estimates.  For example, $f=0.5$ if we require $L_{peak,90}$ instead of $L_{peak,50}$.   
If we interpret this as the shock breaking out of the wind, then the wind density parameter must be
\begin{equation}
      { \dot{M} \over v_w } = {4\pi c r_s \over v_s \kappa_T } \simeq
           = 10^{-3.2} 
          \left( { R(\tau_V=1) \over \tau_{peak} 10^{17}~\hbox{cm} }\right)^2 
            \left( { T_{dest} \over 1500~\hbox{K} }\right)^{5/2}
          \left( { 1000~\hbox{km/s} \over f v_s } \right)^2 
        {M_\odot/\hbox{year} \over \hbox{km/s}},
\end{equation}
which is an order of magnitude larger than the wind densities inferred from the progenitor properties (Fig.~\ref{fig:mwind}).
Thus, the shock breakout radius must correspond to the stellar radius $R_* \simeq r_s$.  This implies a 
stellar temperature of
\begin{equation}  
       T_* = \left( { L_* \over 4 \pi \sigma R_*^2 }  \right)^{1/4} 
      \simeq 4000 \left( { L_* \over 10^5 L_\odot  }\right)^{1/4} 
         \left( { \tau_{peak} 10^{17} \hbox{cm}\over R(\tau_V=1) } \right)
              \left( { 1500~\hbox{K} \over T_{dest} } \right)^{5/4}
              \left( { f v_s \over 1000~\hbox{km/s} }\right)^{1/2}
          ~\hbox{K}.
        \label{eqn:tstar}
\end{equation} 
For our nominal estimates of the progenitor luminosities and wind properties from \S\ref{sec:progenitors} 
and the optical depths at peak from \S\ref{sec:transients}, this yields stellar temperatures of 
order $1400$~K with formal uncertainties of order 30\%, but the estimate is really dominated by systematic 
uncertainties in the models and dust destruction
physics -- temperature is a variable where logarithmic accuracy is not really adequate.  Nonetheless,  
the estimate is far more consistent with a cool, red supergiant as the progenitor, an EAGB star, than with 
a hotter, bluer star.  The break out luminosity increases with the size of the star, $L_s \propto r_s$,
because the larger surface area ($\propto r_s^2$) matters more than the lower shock temperature
($T_s^4 \propto 1/r_s$, Eqn.~\ref{eqn:tbshock}). The duration of the break out peak is of order the light crossing time 
$r_s/c \sim 10^4$~sec or less, so the energy released in the peak is less than that of the observed transient. 

\begin{figure}[t]
\centerline{\includegraphics[width=5.5in]{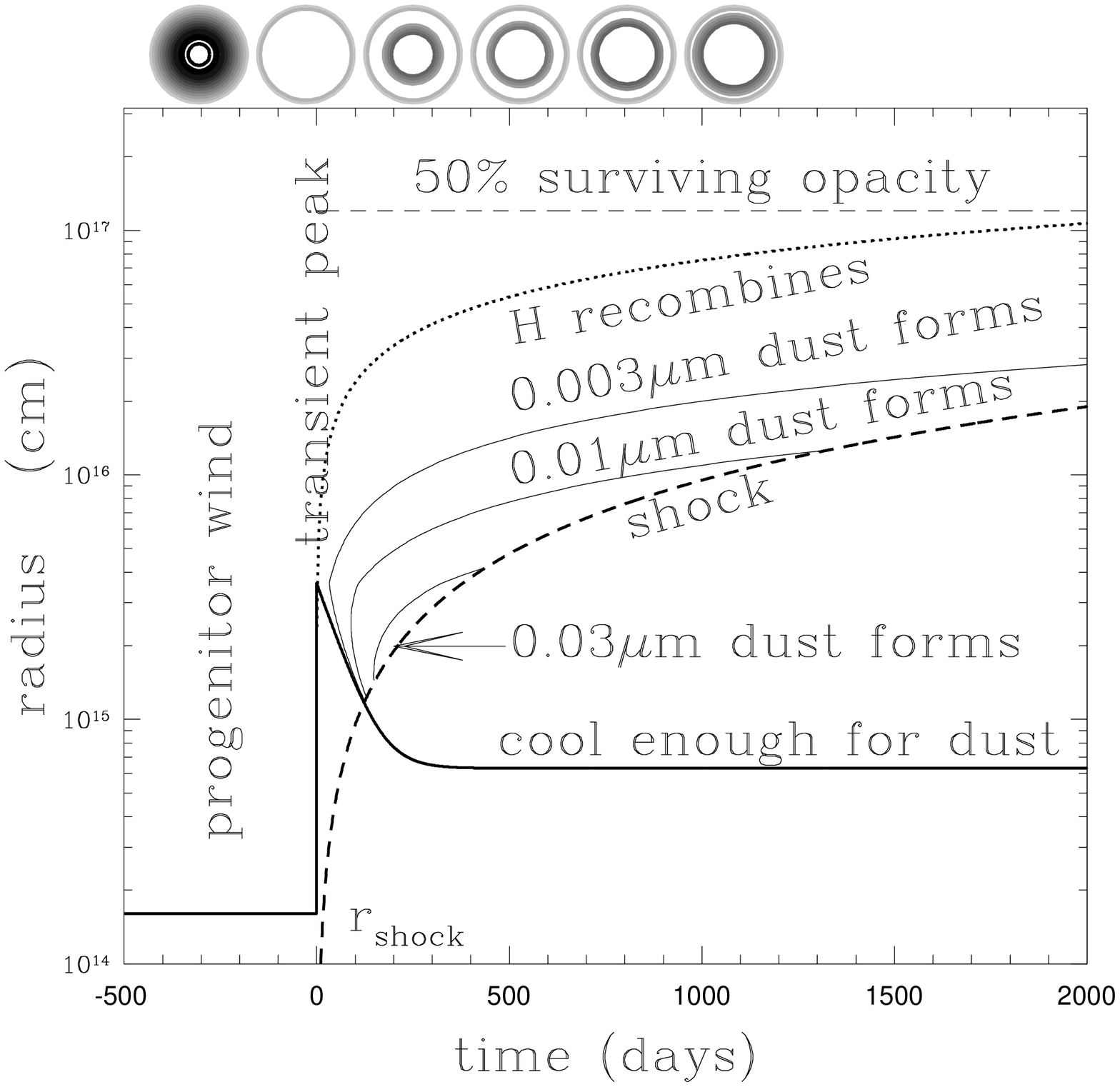}}
\caption{ 
  Physical scales relevant for dust formation as a function of time and scaled for SN~2008S.  
  The heavy solid line shows the radius $R_{form}$
  (Eqn.~\ref{eqn:dform}) interior to which small grains would be destroyed.  The heavy dashed 
  line shows the radius of the outgoing shock.  The horizontal dashed line shows the radius at
  which 50\% of the dust opacity would be destroyed given a shock break out luminosity of
  $4 \times 10^{10} L_\odot$ (Eqn.~\ref{eqn:rd50}).  The dotted shows the radius interior
  to which hydrogen can have recombined.  Finally, the light black lines show the radii
  interior to which grains can have regrown to $0.003\mu$m, $0.01\mu$m, and $0.03\mu$m, 
  respectively.  These would move outwards in radius by a factor of two for the coagulation
  model.  The cartoons at the top roughly illustrate where dust is located and its evolution
  in radius and optical depth.
  }
\label{fig:dhistory}
\end{figure}

\begin{figure}[t]
\centerline{\includegraphics[width=5.5in]{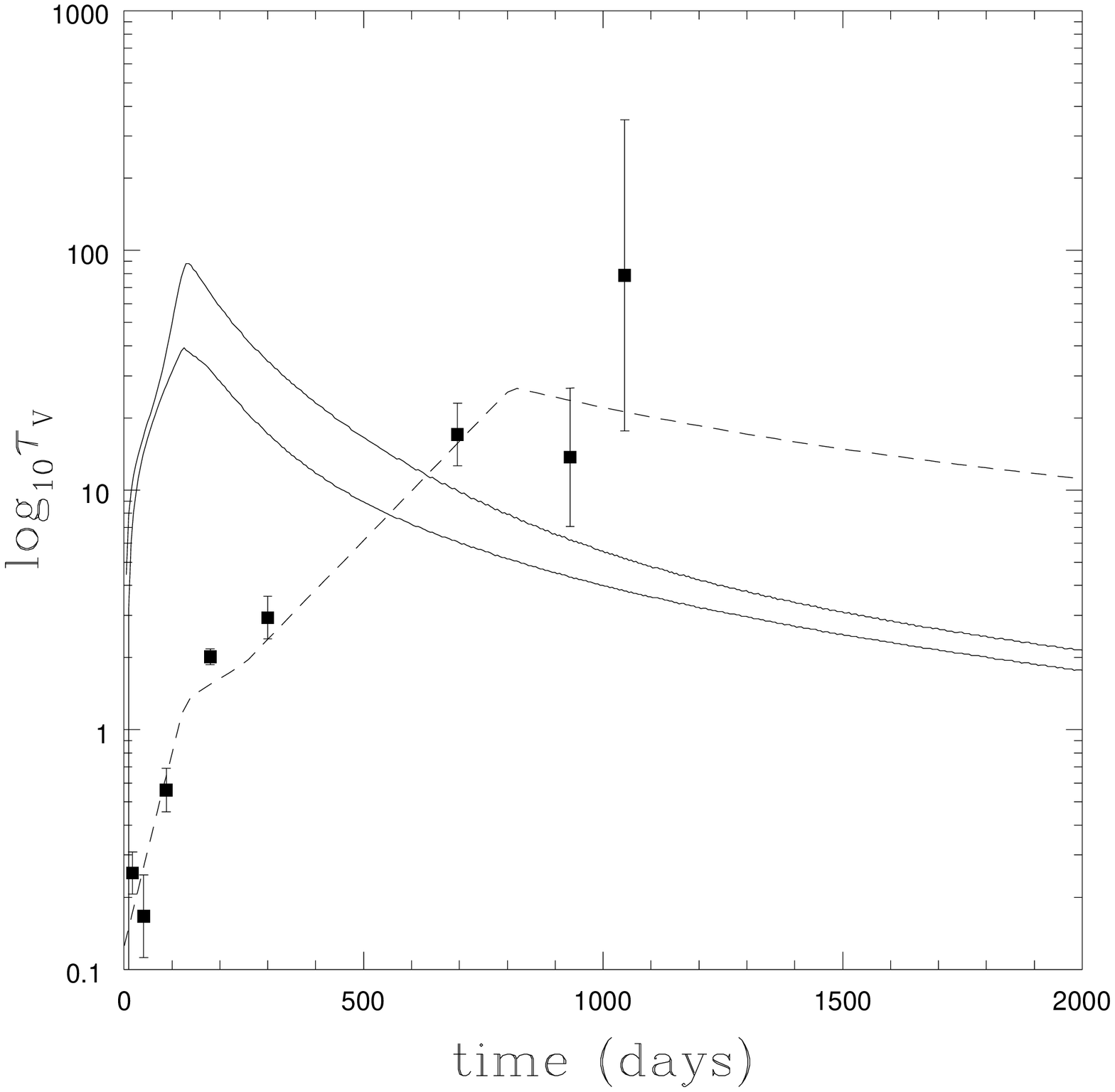}}
\caption{ Simple models for the reformation of dust in the wind of SN~2008S (solid curves).
  The lower curve assumes the growth time scale of Eqn.~\ref{eqn:grow}, and the upper
  curve assumes a rate 4 times faster, corresponding to growth by coagulation rather than
  monomer accretion.  The points and dashed curve show the optical depth estimates from the fits to the
  SEDs and the lightcurves, respectively, as in Fig.~\ref{fig:tevolve}. Given the simplicity 
  of the models, only qualitative agreement is to be expected.  
  }
\label{fig:dusthist}
\end{figure}

\subsection{Recombination, Cooling and Dust Reformation}
\label{sec:reform}

The transient models from \S\ref{sec:transients} appear to show that (1) the dust optical depth rebuilds from the
outside in, (2) that some optical depth returns quickly, but (3) that it takes several years for the optical depth to rebuild
to levels close to that of the progenitor wind.  Sadly we lack the continuous, detailed monitoring of the mid-IR
spectral energy distributions that would have allowed a careful exploration of this process.  In this section
we outline some of the relevant physical issues and  Fig.~\ref{fig:dhistory} outlines our model for the process 

The temperature of the shock break out radiation is high, $T_s \sim 50000$~K (Eqn.~\ref{eqn:tbshock}), so the wind is 
photoionized by the break out radiation.  For pure hydrogen, the minimum ionizing flux needed to maintain the ionization is
\begin{equation}
    Q_0 = { 4 \pi \alpha_B R(\tau_V=1)^2 \over R m_p^2 \kappa_V^2 }   
       \simeq 2 \times 10^{49} \left( { \alpha_B \over 10^{-13}~\hbox{cm}^3/\hbox{s} } \right)
              \left( { 140~\hbox{cm}^2/\hbox{g}  \over \kappa_V } \right)^2
              \left(  { R(\tau_V=1) \over  10^{17} \hbox{cm} } \right) \left( { R(\tau_V=1) \over 10 R } \right)
          ~\hbox{s}^{-1}
\end{equation}
which cannot be supplied by the transients even at their observed peaks (low luminosity, low temperature,
and $R(\tau_V=1)/R \sim 10^3$).  We scale this and the subsequent expressions to the radius $R = 10^{-1} R(\tau_V=1) \simeq 10^{16}$~cm, 
roughly corresponding to the shock radius at the present time ($\sim 3$~years at $1000$~km/s).  
Thus, the wind recombines on time scales (for hydrogen) of  
\begin{equation} 
   t_R = { \mu m_p \over \alpha_B \rho_w  } = { \mu m_p R(\tau_V=1) \kappa_{V} \over \alpha_B } \left( { R \over  R(\tau_V=1) } \right)^2
          \simeq 0.43 \left( { R(\tau_V=1) \over  10^{17} \hbox{cm} } \right) \left( { 10^{-13} \hbox{cm}^3/\hbox{s} \over \alpha_B } \right)
          \left( { 10 R \over  R(\tau_V=1) } \right)^2 ~\hbox{years}
       \label{eqn:recomb}
\end{equation}
that are very short in the interior and only become relatively long as you approach $R(\tau_V=1)$, as
shown in  Fig.~\ref{fig:dhistory}.  While
the transients cannot keep hydrogen and helium in the wind fully ionized due to their low temperatures, 
they have no difficulty keeping sodium photoionized, as discussed in \S\ref{sec:transients}.  At late 
times, as we discuss in \S\ref{sec:xray}, the source of the luminosity is non-thermal emission from a
shock, and ultimately this should reionize the wind because the required number of ionizing photons
steadily diminishes.   

Then as the gas cools, the dust can reform.  The cooling time scale is of order
\begin{equation}
    t_{cool} = { 3 k T \mu m_p \kappa_V r^2 \over 2 \Lambda R(\tau_V=1)}
            \simeq 0.089 \left( { T \over 10^4 ~\hbox{K} } \right) 
                    \left( { 10^{-25} ~\hbox{ergs cm}^3/\hbox{s} \over \Lambda } \right)
               \left( { R(\tau_V=1) \over  10^{17} \hbox{cm} } \right) \left( { 10 R \over  R(\tau_V=1) } \right)^2 ~\hbox{years}
           \label{eqn:tcool}
\end{equation}
where a cooling rate of $ \Lambda \simeq 10^{-25}$~ergs~cm$^3$/s is approximately correct
for $10^3 < T < 10^4$~K (e.g. \citealt{Schure2009}).  Thus, the gas also cools rapidly
after the peak, with the cooling radius being only slightly smaller than the recombination
radius in  Fig.~\ref{fig:dhistory}.   

Once the temperature is low enough, $T \ltorder T_{dest}$ either the dust begins to form anew
following a burst of nucleation in the now supersaturated vapor or any residual grains can
begin to grow.    
Given a steadily declining temperature, nucleation is not a bottleneck, although the radius
for inhibiting dust formation is larger than the dust destruction radius because less flux
is needed to destroy the smallest grains.  For a $7000$~K transient and $T_{dest}=1500$~K, 
graphitic dust can begin to form outside the radius 
\begin{equation}
         R_{form} \simeq 10^{15.9} \left( { L \over 10^8 L_\odot } \right)^{1/2}~\hbox{cm},
       \label{eqn:dform}
\end{equation}
estimated from the heating of the smallest grains.  We used this estimate of the destruction
radius in Fig.~\ref{fig:revolve} to illustrate where dust can begin to reform.  For both
transients, $R_{form}$ is outside the shock radius only for the first $\sim 100$~days,
as shown in  Fig.~\ref{fig:dhistory} based on the exponential plus constant luminosity 
models for the transients from \S\ref{sec:transients}.  For comparison,  Fig.~\ref{fig:dhistory} also shows
the radius at which 50\% of the dust opacity will survive a $L_{peak} = 4 \times 10^{10}L_\odot$ 
break out shock based on Eqn.~\ref{eqn:rd50}.   

The bottleneck for reforming the dust is the growth rate of the particles after nucleation, where 
the time scale to grow a (spherical) grain of radius $a$ by adding monomers assuming a sticking 
fraction of unity is of order
\begin{equation}
      t_{grow} \simeq { 4 \rho_{bulk} a \over  v_c X_c \rho }
         \simeq 19 \left( { a \over 0.1 \mu\hbox{m} } \right) \left( { 0.02 \over X_c }\right)
           \left( { R(\tau_V=1) \over  10^{17} \hbox{cm} } \right) \left( { 10 R \over  R(\tau_V=1) } \right)^2 ~\hbox{years}
           \label{eqn:grow}
\end{equation}  
where $v_c \simeq v_w \simeq 10$~km/s is the collision velocity, and $X_c$ is the carbon mass fraction
in the wind (we have used a relatively high value appropriate for a carbon rich EAGB wind (\citealt{Abia2000})).
If the particles are growing by coagulation rather than monomer accretion, the growth
rates are a four times faster, but significantly shorter growth times could only come from incorporating
significant numbers of the far more abundant hydrogen atoms, having a reduced dimensionality (e.g.
sheets) or density, or starting from more and larger surviving particles at larger radii.  In practice, 
the latter effect is certainly present, so Eqn.~\ref{eqn:grow} does exaggerate the difficulty of reforming
dust at larger radii.  We should also note that there is no need to fully repopulate the $0.005<a<0.25\mu$m
size range of the standard DUSTY \cite{Mathis1977} models to recover most of the opacity.  
DUSTY models using distributions with maximum grain sizes of $0.01$, $0.025$, $0.05$ or $0.1\mu$m 
instead of $0.25\mu$m have visual opacities smaller only 
by factors of $3.4$, $3.4$, $2.9$ and $1.6$, respectively, consistent with the rapid return of a
significant fraction of the pre-transient opacity followed by a slower rise towards the total pre-transient opacity.  

Fig.~\ref{fig:dhistory} illustrates the cycle of dust formation for SN~2008S.  Assuming growth begins after
both recombination is completed and the radiative heating is sufficiently low, dust begins to
reform at an intermediate radius where the formation time scale is relatively long.  However,
since the cooling and growth time scales are significantly shorter at smaller radii, the 
optical-depth weighted radius of the dust will rapidly shrink, roughly tracking $R_{form}$,
until it reaches the expanding radius of the shock.  Our models essentially represent the
evolution of this optical-depth weighted radius.  A variant of the models where we used
two shells, one at large radius to represent the dust surviving the explosion, and one 
expanding with the shock, could not reproduce the light curves as well.  We attempt an
explicit model of the evolution of the optical depth in Fig.~\ref{fig:dusthist}.  In
this model, the dust begins to reform at each radius after the longer of either the
recombination time (Eqn.~\ref{eqn:recomb}) or the time needed to lie outside the region
radiatively heated to $T_{dest}$ (Eqn.~\ref{eqn:dform}).  The maximum particle size $a_{max}$
is then determined from Eqn.~\ref{eqn:grow}, and the opacity is found by the scaling of the
DUSTY model opacities with $a_{max}$.  Qualitatively this reproduces the behavior of
the optical depth in the models, which are also shown in Fig.~\ref{fig:dhistory}. Quantitatively,
the optical depth rises too quickly and is somewhat low at the time of our last SED model.
The early time differences could be due to oversimplifying the temperature of the dust at
smaller radii where there may be additional heating and destruction  mechanisms associated 
with the ionizing radiation from the expanding shock front (see \S\ref{sec:xray}).  Overall,
however, the physical characteristics of the winds are broadly consistent with reforming
the necessary amounts of dust.  

\begin{figure}[t]
\centerline{\includegraphics[width=5.5in]{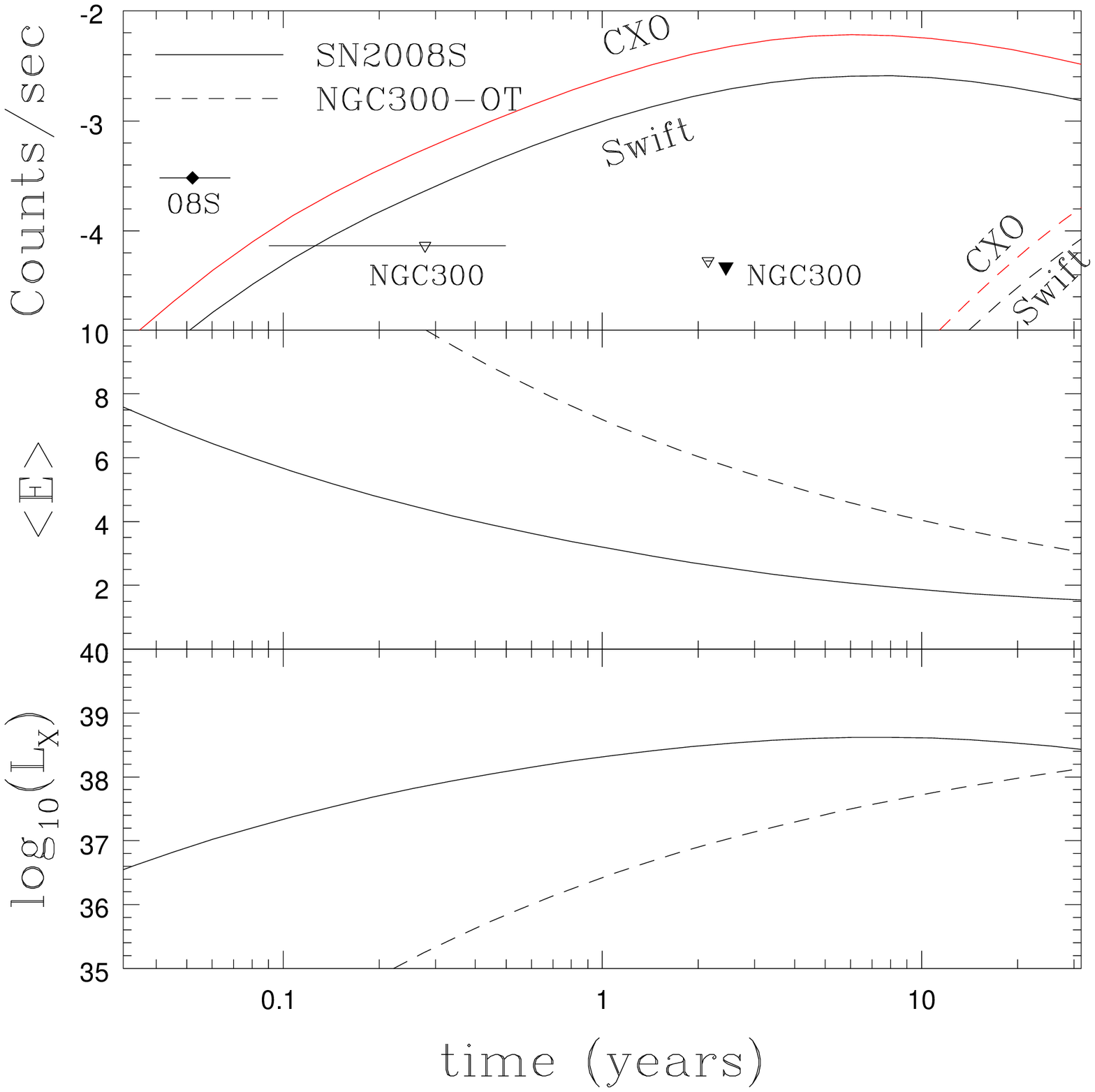}}
\caption{ The predicted X-ray properties of SN~2008S (solid) and the NGC~300-OT (dashed).  The
  bottom panel shows the estimated luminosity including both absorption and the crude estimate
  of the balance between radiation and expansion.  The middle panel shows the mean X-ray 
  energy after absorption.  The top panel shows estimates of the count rates for SWIFT
  (lower count rates, black) and Chandra (higher count rates, red).  The points show the
  count rates for finding 3 counts in archival SWIFT (black diamond for SN~2008S open
  triangles for the NGC~300-OT) and Chandra (filled triangle for the NGC~300-OT) 
  observations given their integration times.  Multiple SWIFT epochs were combined
  into a single limit. 
  }
\label{fig:xray}
\end{figure}

\subsection{X-ray Emission from the Shocks Powers the Late-Time Light Curves}
\label{sec:xray}

Even after almost three years, these sources are still an order of magnitude more luminous than their progenitors, and
the present emission is most likely due to shock heating the dense wind.  For a shock wave propagating outward at
velocity $v_s >> v_w$, the shock releases energy (e.g. \citealt{Chevalier1982}, \citealt{Fransson1982}, 
\citealt{Chugai1992}, \citealt{Chugai1994}, \citealt{Chevalier1994})
\begin{equation} 
   L_S \simeq { \dot{M} \over 2 } { v_s^3 \over v_w } = { 2\pi R(\tau_V=1) v_s^3 \over \kappa_{V} }
     = 1.2 \times 10^6 \left( { v_s \over 1000 \hbox{km/s} } \right)^3 \left( {  R(\tau_V=1) \over 10^{17}~\hbox{cm} } \right)
                  \left({ 140~\hbox{cm}^2/\hbox{g} \over \kappa_{V} }\right) L_\odot
   \label{eqn:lshock}
\end{equation}
although this energy can be lost due to expansion rather than radiated.  In detail, there is both a forward and a reverse shock, separated
by a contact discontinuity. Models of X-ray emission for typical SN focus on the reverse shock because
of its lower temperature.  Here, the low velocities make the forward shock a source of very soft, readily 
absorbed emission as well.  Exactly what should be used for $v_s$ is unclear, since the line widths
in normal Type~IIn are not believed to be representative of the true shock velocity, but as with the light
curve models in \S\ref{sec:transients} we will scale results based on the FWHM estimates from \cite{Smith2011} 
of $v_s=1100$~km/s and $560$~km/s for SN~2008S and the NGC~300-OT, respectively.  Because
of the low shock velocities, the post-shock temperatures are low
\begin{equation}
      E_s = { 3 \mu \over 16 } m_p v_s^2 = 1.2 \left( { v_s \over 1000 \hbox{km/s} }\right)^2 ~\hbox{keV}
        \label{eqn:eshock}
\end{equation}
for a mean molecular weight $\mu=0.6$.
The optical depth of the swept up wind is the same as that of the wind exterior to the shock for the same opacity, 
so the swept up wind material quickly becomes Thomson optically thin,
\begin{equation}
     \tau_T \simeq { R(\tau_V=1) \over R } {\kappa_T \over \kappa_V }
            \simeq 0.09 \left( { R(\tau_V=1) \over 10^{17}~\hbox{cm} } \right) \left( { 1000~\hbox{km/s} \over v_s } \right)
                  \left( { \hbox{year} \over t }\right).
          \label{eqn:thomson}
\end{equation}   
The low Thomson optical depth of the wind essentially corresponds to the conclusion in \S\ref{sec:destroy} that
the wind density is too low to be the source of the shock breakout luminosity.  
Typical models of these shocks consider cooling by adiabatic expansion, Compton cooling and free-free
emission, but generally consider much faster shocks ($v_s \sim 10^4$~km/s) in less dense winds, finding that
the early-time cooling is dominated by Compton cooling.  In these systems, free-free emission dominates
over Compton cooling, with a cooling rate of order $ \Lambda = A_0 n^2 (T/T_0)^{1/2}$ for $T_0 = 1~\hbox{keV} \simeq 10^7$~K
and $A_0 \simeq 2 \times 10^{-24}$~ergs~cm$^3$/s (e.g. \citealt{Rybicki1979}), so
the ratio of the cooling time, $t_{cool}=3 n k T_s/2 \Lambda$, to the expansion time, $t_{exp}=R/v_s$, is
\begin{equation} {t_{cool} \over t_{exp}} 
     \simeq 0.5\left( { v_s \over 1000~\hbox{km/s}} \right)^2 \left( { \kappa_{V} \over 140~\hbox{cm}^2/\hbox{g} }\right) 
        \left( { 4 \over C_{comp} } \right) \left( { 10 R \over R(\tau_V=1) }\right) 
        \label{eqn:balance}
\end{equation}
where $C_{comp}>(\Gamma+1)/(\Gamma-1)=4$ is the ratio of the post-shock density to the pre-shock density.
Thus, the shock energy will be efficiently radiated as soft X-rays until the shock radius approaches $R(\tau_V=1)$. 

Unlike typical SN, the combination of high wind densities and a very soft X-ray spectrum mean that 
the X-ray emission is fully absorbed by the unshocked wind material and radiated in the mid-IR by
the reformed dust while the wind is optically thick.  Following   \cite{Draine1991}, we model the 
(bound-free) X-ray opacity of the wind as
\begin{eqnarray} 
   \kappa_H(E) &= &17 (E/\hbox{keV})^{-3}~\hbox{cm}^2/\hbox{g} \\
   \kappa_Z(E) &= &85 \left( 1+ (E/\hbox{keV})^{2.5}\right)^{-1}~\hbox{cm}^2/\hbox{g} \nonumber
        \label{eqn:kapx}
\end{eqnarray} 
from neutral hydrogen/helium and heavier elements, respectively.  This implies X-ray optical depths of
\begin{eqnarray} 
   \tau_H(E) &= &1.2 \left( { 140\hbox{cm}^2/\hbox{g} \over \kappa_{V}} \right)
          \left( { R(\tau_V=1) \over 10 R }\right) \left( { E \over \hbox{keV}}\right)^{-3} \\
     \label{eqn:taux}
   \tau_Z(E) &= &6.1 \left( { 140\hbox{cm}^2/\hbox{g} \over \kappa_{V}} \right)
          \left( { R(\tau_V=1) \over 10 R }\right) { 1 \over 1+ (E/\hbox{keV})^{2.5} }. \nonumber
\end{eqnarray} 
from radius $R$ to infinity.  The corrections for the material outside the recombination radius 
(Eqn.~\ref{eqn:recomb}) are not important until very late.  The free-free emission spectrum is
\begin{equation}   
    F_\nu \propto g(E/E_0) \exp(-E/E_0) 
\end{equation}
where  $g(x)$ 
is the Gaunt factor. For an X-ray temperature of $E_0=1$~keV,  the fraction of the X-ray energy 
escaping the wind for $R/R(\tau_V=1)=0.01$, $0.1$, and $1$ is only $0.2\%$, $5\%$, and $27\%$,
with mean energies of $4.8$, $2.6$ and $1.5$~keV.  Slightly more escapes for SN~2008S ($E_0\simeq 1.2$~keV)
and quite a bit less escapes the NGC~300-OT ($E_0 \simeq 0.4$~keV).  While soft X-ray absorption from
SN shocks has been discussed in the context of normal SN models (e.g. \citealt{Fransson1982}), the overall
spectrum is assumed to be much harder (because $v_s$ is larger), so only small fractions of the total luminosity are absorbed.

We can approximate the expected X-ray emission as
\begin{equation}
          L_X \simeq  L_S { 1-f_{abs}  \over 1 + t_{exp}/t_{cool} }
        \label{eqn:lumx}
\end{equation}
where $L_S$ is the shock luminosity from Eqn.~\ref{eqn:lshock}, $1-f_{abs}$ is the unabsorbed fraction estimated
by combining the wind model, the X-ray optical depths in Eqn.~\ref{eqn:taux} and the shock energies from 
Eqn.~\ref{eqn:eshock}, and the denominator approximates the balance between radiative
and expansion losses from Eqn.~\ref{eqn:balance}.  When the cooling time is short, all the energy is radiated, while when it is long,
only the fraction $t_{cool}/t_{exp}$ is radiated and the remainder is lost to adiabatic expansion.
The evolutions of the expected X-ray luminosity and the mean photon energy for the two systems are shown
in Fig.~\ref{fig:xray}.   Without absorption, both systems should be bright X-ray sources.  With 
absorption, however, they are relative faint, hard X-ray sources in the early phases, and it is not
at all surprising that they were not detected in the early SWIFT observations 
(\citealt{Botticella2009}, \citealt{Berger2009}).  While SN~2008S has not been reobserved, 
the NGC~300-OT was reobserved after roughly two years by both SWIFT (\citealt{Berger2010})
and Chandra (CXO observation ID 12238/PI: Williams).  Fig.~\ref{fig:xray} indicates the 
count rates at which these various observations would detect 3 source photons, where
we estimated the count rates using the Chandra/SWIFT exposure time calculators and 
absorbed thermal bremsstrahlung models based on the gas column densities in our nominal wind models for
the two sources.  Neither source should have been detectable in any of the observations to
date.  SN~2008S should now be relatively easy to detect, with count rates of order $10^{-3}$/sec,
but has not been observed, while the NGC~300-OT is essentially undetectable -- 
it is less luminous, softer, and more heavily obscured.  Bear in mind, however, that at 
fixed shock luminosity, the detectability of the X-ray emission depends exponentially on the 
shock velocity because the X-ray optical depth of the wind scales as $L_s/v_s^4$ 
(three powers of $v_s$ from the shock luminosity, Eqn.~\ref{eqn:lshock}, and one
power because of the slower expansion through the wind), and the typical energy scales 
as $v_s^2$ (Eqn.~\ref{eqn:eshock}).  Small reductions in $v_s$ dramatically increase the
overall absorption.

Such shocks are also expected to produce radio emission, which has not been observed
(\citealt{Chandra2008}, \citealt{Berger2009}), and this is also used as evidence
against an explosive transient.  In fact, no early time radio emission is
likely to escape the wind exterior to the shock because absorbing the X-ray emission
will create an ionized layer in the unshocked wind exterior to the shock.
If this ionized layer has a thickness $\Delta R$, and electron temperature 
$T_e \simeq 10^4$~K, then the free-free optical depth of the layer is 
\begin{equation} 
   \tau_\nu \sim 2000 \nu_{GHz}^{-2.1}\left( { 10^4~\hbox{K} \over T_e  }\right)^{1.35}
          \left( { 10 \Delta R \over R } \right) \left( { 10^{17}~\hbox{cm} \over R(\tau_V=1) } \right)
        \left( { R(\tau_V=1) \over 10 R }\right)^3
\end{equation}
and these systems will be very optically thick to GHz radio emission in their early phases
and should not have been detectable.  The time scale for the optical depth to fall to $\tau_\nu=1$ is
\begin{equation}
    t_\nu \sim 40 \nu_{GHz}^{-0.7} \left( { R(\tau_V=1) \over 10^{17}~\hbox{cm} } \right)^{2/3}
            \left( {10 \Delta R \over R } \right)^{1/3}
             \left( { 10^4~\hbox{K} \over T_e } \right)^{0.45} 
             \left( { 1000~\hbox{km/s} \over v_s } \right) ~\hbox{years}
\end{equation}
ignoring any other potential source of opacity at these wavelengths.  Thus, at late times
these systems may be detectable in the radio.  More generally radio emission from
such high density winds or from shocked, massive shells of material will be very different from 
the significantly lower regimes of circumstellar densities that are usually considered for
radio supernovae (e.g. \citealt{Chevalier2006}). 

\subsection{Mid-IR, near-IR and Optical Emission}
\label{sec:optir}

The current mid-IR emission is simply the absorbed X-ray flux combined with any emission 
from dust heated and destroyed by the shock.  The absorbed X-ray emission leads to a 
mid-IR luminosity of 
\begin{equation}
          L_{IR} \simeq  L_S { f_{abs}  \over 1 + t_{exp}/t_{cool} }.
\end{equation}
which is simply the absorbed counterpart to Eqn.~\ref{eqn:lumx}, as long as the dust
optical depth remains high.  Here we are ignoring the extra emission from dust directly 
heated by the shock, which typically radiates $\sim 10\%$ of the shock luminosity 
(\citealt{Draine1981}).  The dust temperature is simply determined by the available
luminosity, since emission by the dust is the only means of
radiating the absorbed energy.  Thus, the
dust temperature must be or order
\begin{equation}
      T_d \simeq \left( { L_{IR} \over 16 \pi \sigma r_d^2 }\right)^{1/4}
          = 440 \left( { L_{IR} \over 10^6 L_\odot } \right)^{1/4} \left( { 1000~\hbox{km/s} \over v_s }\right)^{1/2}
                  \left( { 1000~\hbox{days} \over t } \right)^{1/2}~\hbox{K}.
\end{equation}
where $r_{d} $ is the dust radius.  Driven by the declining luminosity and the increasing
dust radius, the mid-IR emission slowly declines and shifts to longer wavelengths, as
we see in Figs.~\ref{fig:lc1} and \ref{fig:lc2}. 
This phase lasts until the shock expands to the point that the optical depth of the reformed dust 
begins to clear.  If all the dust has reformed,  it takes 
\begin{equation}
      t (\tau_\lambda) = { R(\tau_V=1) \over v_s \tau_\lambda } {\kappa_\lambda \over \kappa_V }
             = 31  \left( {  R(\tau_V=1) \over 10^{17}~\hbox{cm} } \right) 
                       \left( { 1000~\hbox{km/s} \over v_s } \right)
                      { \kappa_\lambda \over \kappa_V } { 1 \over \tau_\lambda}~\hbox{years}
\end{equation}
for the optical depth down to the shock to be $\tau_\lambda$.  In Figs.~\ref{fig:lc1} and
\ref{fig:lc2} this leads to the sequential recovery of the near-IR and then the optical fluxes, with the
present period being the phase where the absorption is worst.  In these models, the near-IR
fluxes should begin to recover in the immediate future.

\begin{figure}[p]
\centerline{\includegraphics[width=5.5in]{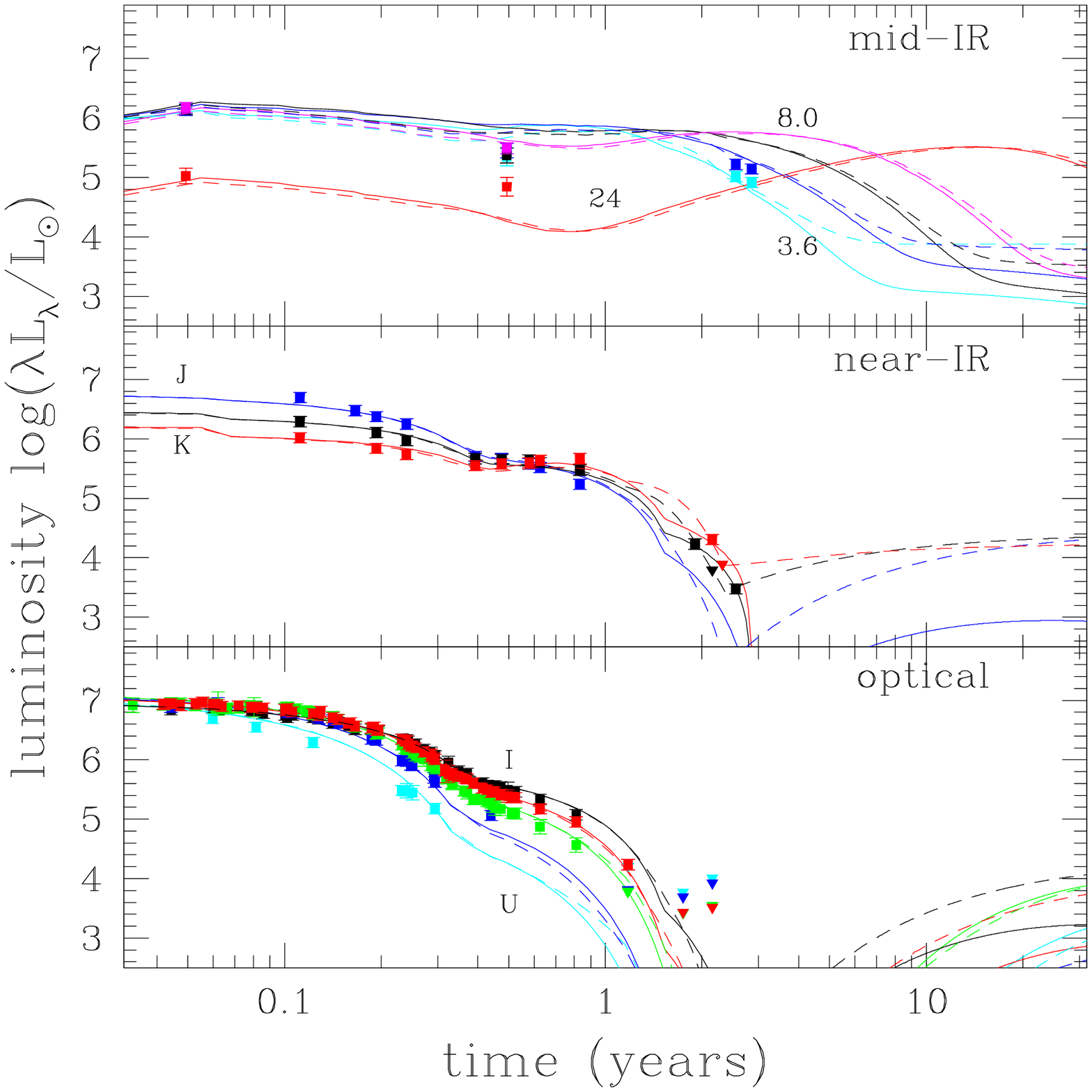}}
\caption{ Alternate light curve models for SN~2008S.  As in Fig.~\ref{fig:lc1},
  the top, middle and bottom panels show the mid-IR (IRAC $3.6$, $4.5$, $5.8$, $8.0$ 
  and MIPS $24\mu$m), near-IR (J, H and K), and optical (UBVRI) light curves.
  The solid lines show the light curves using an emission line model for the
  shock spectrum, and the dashed lines add an unobscured $3000$~K black-body 
  with the luminosity of the progenitor.  As expected, the differences between
  the models appear in the near-IR and optical as the optical depth to the 
  shock front diminishes.  The shock model has relatively little near-IR
  emission, so the near-IR emission never recovers, while the shock plus
  progenitor model is brighter in the near-IR than the original model
  because of the cooler progenitor black body temperature.  The latter model assumes
  that there is no significant dust optical depth interior to the shock. 
  }
\label{fig:lc1b}
\end{figure}

Unfortunately, the late time details of the light curves in  Figs.~\ref{fig:lc1} and \ref{fig:lc2} are incorrect because
the models simply continue to assign the transient a black body SED of fixed temperature.  This approximation
is adequate when the optical emission is dominated by the transient rather than the shock or when the
emission is all absorbed and re-radiated in the mid-IR, but it is not correct when the dust opacities
become low enough for the line emission from the shock to escape.
The SED of the shock will be a complex line spectrum (e.g. \citealt{Fesen1994} for SN~1980K) extending from the UV
into the mid-IR (e.g. \citealt{Allen2008}) rather than a black body.  The \cite{Allen2008} shock models
do not reach the wind densities needed for these systems, but to the extent we can extrapolate from
the highest density models in their survey, the dominant emission is in the UV (Ly$\alpha$), followed by the 
optical (H$\alpha$) and mid-IR, with relatively little emission in the near-IR.  Hence, the extrapolations
shown in our light curve models are probably over-estimating the recovery in the near-IR.
Fig.~\ref{fig:lc1b} shows the effect on the light curve of SN~2008S if we model the shock emission
using the emission lines and line ratios of the highest density ($n=10^3$~cm$^{-3}$), solar metallicity,
 $v_s=1000$~km/s models from \cite{Allen2008}.  The true density is much higher, and so the
actual line ratios will be strongly affected by the collisional de-excitation effects that
limited their survey of shock emission spectra to $n\leq 10^3$, but it illustrates the 
potential differences from the model used in Fig.~\ref{fig:lc1}.  As a result,
these shock models have no significant recovery of the near-IR flux as the optical depth drops.

\subsection{Did the Stars Survive?}
\label{sec:survivor}

Because the shock luminosity is significantly greater than that of the progenitor, we can
only indirectly constrain the existence of a surviving star.  One probe of this question
is the mass of the ejecta.  Using the expansion velocities of $v_s=1100$ and $560$~km/s from
\cite{Smith2011} for the transients, and assuming the radiated energy estimates from
\S\ref{sec:transients} are fraction $f$ of the kinetic energy, the ejected masses are of order 
$M_e \sim 0.02 f^{-1} M_\odot$ and $0.25f ^{-1} M_\odot$ for SN~2008S and the NGC~300-OT
transient.  Since destroying the dust required an explosive 
transient, we would expect $ f \sim 0.01$ as is typical of supernovae 
rather than $ f \sim 1$ similar to $\eta$ Carinae (e.g. \citealt{Smith2011}), but this argument is too crude
to determine the survival of the star.  The roughly constant shock luminosity 
means that the ejected mass must significantly exceed the swept up wind mass,
so $M_e \gtorder 3 \dot{M} v_s t/v_w$, which after $t\simeq 3$~years corresponds
to $M_e \gtorder 0.05 M_\odot$ and $0.25 M_\odot$ for our standard parameters,
weakly requiring $f \gtorder 1$.  Unfortunately, the time scales for estimating
$M_e$ from the evolution of the transients are long.    

This leaves the problem of directly detecting a surviving star.  As the shock expands, 
the pre-existing wind will stop cloaking the progenitor and the shock luminosity will 
fade due to the decreasing radiative efficiency and will likely be dominated by X-ray 
and line emissions that are not characteristic of the progenitor.  Fig.~\ref{fig:lc1b}
shows the effect of including the SN~2008S progenitor as a $T_*=3000$~K black body
in the shock model for the late-time emissions.  This assumes that no dust has formed
in or behind the shock but includes the obscuration from the wind exterior to the 
shock.  Obviously, if dust forms either near the contact discontinuity between the
ejecta and the wind or in the ejecta, there is easily enough material to fully
cloak the star.  Moreover, the luminosity of the shell plus any surviving star is 
insufficient to maintain a temperature above the dust destruction temperature
in the region interior to the reverse shock.      

\section{Summary}
\label{sec:conclusions}

The available data for the SN~2008S and the NGC~300-OT are consistent with a relatively
simple, self-consistent physical picture of the systems.  The progenitor stars are luminous,
cool, red supergiants, most likely EAGB stars based on their positions in mid-IR CMDs 
in a short-lived state with very high mass loss rates  
(\citealt{Thompson2009}, \citealt{Khan2010}).
The very dense winds form so much
dust that they are optically thick in the mid-IR with dust photospheres on scales of a
few $10^{15}$~cm and optical photospheres near $10^{17}$~cm.  They are best modeled with
graphitic dusts (\citealt{Prieto2009}, \citealt{Wesson2010}), and the radiatively driven
dusty wind models of \cite{Ivezic2010} can produce such winds at the $\sim 9M_\odot$ 
mass scale of EAGB stars assuming graphitic, but not silicate, dusts.

Both stars then underwent an explosive transient with a short duration, very high 
luminosity shock breakout spike.  This unobserved luminosity peak is necessary because
the observed luminosity peaks are not nearly bright enough
to destroy dust to the distances needed to have the transient
peaks little extincted.  We know from early-time mid-IR observations of \cite{Wesson2010} 
that SN~2008S had a modest surviving dust optical depth, and our models of the early-time UV
through near-IR SED of the NGC~300-OT from \cite{Berger2009} and \cite{Bond2009} imply 
a surviving visual optical depth of a few, which is consistent with the Balmer 
decrement observed by \cite{Berger2009}.   \cite{Berger2009} argue that the weak
Na~I~D absorption in spectra of the NGC~300-OT imply no surviving dust, but we
argue that sodium must remain photoionized independent of any surviving dust 
and would be uncorrelated with any surviving
dust if present.  Similarly, the dense circumstellar wind
is initially opaque to X-rays and requires very little ionization to be opaque to
radio emission, so the failure to detect the systems at both X-ray and radio wavelengths
at early-times is also a natural consequence of the progenitor properties even
for an explosive transient.  The
stellar radius required by the shock break out model also requires a large, cool
star, again consistent with the EAGB picture for the progenitor stars.

The evolution of the transient SEDs with time is largely controlled by the reformation
of the dust in the progenitor wind outside the expanding transient shock wave.  The
average dust radius appears to move inwards from near the original optical photosphere as the dust
destruction radius shrinks.  The dust radius then appears to begin to expand again to
track the estimated radius of the shock wave passing through the wind, although the
evidence for this is less striking than the initial shrinking of the dust radius.  
Initially, the optical depth is significantly less than
that of the progenitor wind outside the same radius, but the fraction of the dust that
reforms steadily rises, with most of the dust having reformed after about $2$-$3$ years.
The dust radius shrinks faster than the transient luminosity drops, so that the dust
temperature peaks 8--9 months after the transient peak, and this rising dust temperature balances
the falling luminosity to produce relatively extended, slowly varying near-IR light curves.
The optical fluxes drop very rapidly from the combination of the dropping luminosity
and the rising optical depths.  Although the luminosity eventually becomes roughly 
constant, the optical depth continues to rise and the systems fade in the near-IR.  At present,
neither system is detected at optical wavelengths (\citealt{Prieto2010a}, \citealt{Prieto2011}), 
and SN~2008S is at best marginally
detectable in the near-IR even with HST (\citealt{Prieto2011}).   In the mid-IR, however, both systems
are easily detected with significantly higher temperatures and luminosities than
the progenitor stars.

The average dust radius in our models is an optical depth weighted mean radius, and 
the true evolution is more complicated.  Models with a surviving shell of dust
and a shell of forming at the the expanding shock radius work poorly -- successful
models require dust reforming in the exterior wind and are incompatible with a shell
ejection scenario, just as was true of the pre-transient mid-IR variability. 
The optical depth of the reformed dust is
dominated by the smallest radius at which dust can reform because the
particle growth rates are proportional to density, so the effective dust radius
shrinks rapidly, tracking the decaying luminosity with a delay, until it encounters
the expanding shock.  The simple model of a smoothly shrinking dust radius elides over
a gap at $10^{16}$ to $10^{17}$~cm, between the surviving and reforming dust, where 
the particle growth times are too long for rapid reformation.    
This pattern of a receding dust radius is also reported for the Type~IIP SN~2007it (\citealt{Andrews2011})
and SN~2007od (\citealt{Andrews2010}), although the implied optical depths are much lower.  
Re-forming CSM dust sufficiently quickly requires the high densities of
slow, high $\dot{M}$ winds or massive shells of ejecta.
In most systems, only the normal scenarios are relevant because the particle
growth rates are too slow: dust must either survive the explosion, form in cooled shocked CSM near 
the contact discontinuity or form in the un-reverse-shocked ejecta.  The high CSM densities
that allow rapid dust reformation are yet another new (or odd) property 
of these transients.

The present day, roughly constant luminosities closely match the luminosity expected
from expanding shocks heating the progenitor winds.  The shocks are Thomson optically
thin but radiatively efficient, leading to high $ \sim 10^{5.5} L_\odot$, soft ($0.5$-$1.0$~keV)
X-ray luminosities.  Soft X-ray opacities are, however, not very different from dust
opacities, so the dense progenitor winds simply absorb the shock luminosity and
the energy is ultimately radiated in the mid-IR, leading to the present-day 
spectral energy distributions.  Neither source should have been detectable in
the X-ray observations of these systems to date, despite having intrinsic X-ray
luminosities of order $10^{39}$~ergs/s. As the shock moves outward, the X-rays begin
to escape and should show a spectrum with strong absorption at 0.5-2.0~keV that
will be a powerful probe of the wind column density.  SN~2008S, which has
a lower density progenitor wind and a faster shock velocity,
should now be relatively easy to detect as an X-ray source, while the NGC~300-OT
should still be very difficult to detect because of its higher density wind
and slower shock velocity.  Slower shock velocities reduce the X-ray detectability
partly because the slower shock expansion rate slows the drop in the exterior
wind column density and partly because the lower shock temperature leads to a softer
more easily absorbed X-ray spectrum.  The detectability does, however, depend
exponentially on the assumed shock velocity.  The peak X-ray emission is a balance
between the diminishing absorption column density and the diminishing radiative
efficiency of the shock, peaking on time scales of order 10 years for SN~2008S
and much later for the more slowly evolving, more heavily obscured NGC~300-OT.   
Absorbing the X-ray emission will produce an ionized region in the unshocked
wind just outside the shock, and the resulting free-free opacity will
also block any GHz radio emission produced by the shock for an extended period
of time. Thus, the failure to detect early-time radio emission is also expected.

The balance between dust reformation and shock expansion means that the visual 
optical depth to the shock peaked after roughly $2$-$3$ years and should now be 
dropping linearly with time.  The mid-IR luminosity should steadily diminish,
because there is less absorption, and shift to longer wavelengths, because
the larger dust radius and diminished luminosity lead to lower dust temperatures.
It may be difficult to cleanly separate these trends with only $3.6$ and
$4.5\mu$m warm Spitzer observations.  Near-IR observations will shortly 
be able to penetrate to the shock front, although the visual optical depth to
the shock front will not approach unity for over a decade.  The optical/near-IR
emission from the shock region should be a combination of free-free emission, 
line emission driven by X-ray absorption and shock heated dust.  In the
evolution to date, the luminosity was dominated by the shock emission while 
the dust optical depths are high enough to allow us to ignore these details,
which will not be true of these later phases.  Unfortunately, until the 
luminosity fades or we can penetrate the dust, it is impossible to determine
the fate of the progenitor star.

\acknowledgements 
The author would like to thank J.F.~Beacom, J.A.~Johnson, M. Pinsonneault, J.-L. Prieto, 
R.~Pogge, K.~Sellgren, K.Z.~Stanek, D.~Szczygiel, T.A~Thompson and B.E.~Wyslouzil for extensive discussions.
CSK is supported by NSF grant AST-0908816.
This research has made use of the NASA/IPAC Extragalactic Database (NED) which is operated by the 
Jet Propulsion Laboratory, California Institute of Technology, under contract with the National 
Aeronautics and Space Administration. 

\appendix
\section{Light Curve Fits}

Here we spell out the details of the light curve fitting procedure. We
specify the luminosity and optical depth on a grid of times $t_i$ with
$i=1 \cdots N$ and $t_1\equiv 0$.  The transient luminosity $L_i = L(t_i)$ 
and temperature $T_i(t_i)$ are determined at each grid point, and the value 
at time $t$ is found by
linearly interpolating $\log L_i$ between bracketing times $t_i < t < t_{i+1}$
or fixed to $L_N$/$T_N$ for $t > t_N$.  The visual optical depth $\tau_{V,i}$
is determined at grid points $i=1 \cdots (N-1)$, but fixed to the 
optical depth of the progenitor wind outside the current dust radius
\begin{equation}
  \tau_V(r_d) = { R(\tau_V=1) \over r_d (t)}
\end{equation} 
for $t>t_N$, where the dust radius declines exponentially starting from radius $r_0$ on time scale $t_0$, 
and then begins to expand outwards at the shock velocity $v_s$,
\begin{equation}
   r_d(t) = r_0 \exp(-t/t_0) + v_s t.
\end{equation}
Thus, the last optical depth grid point is $\tau_{V,N}=\tau_V(r_d(t_N))$ and
represents the point where the optical depth of the wind outside the shock
has fully reformed.  The parameters of the model are then the $N$ luminosities
$L_i$ and temperatures $T_i$, the $N-1$ optical depth points $\tau_{V,i}$ ($i=1\cdots N-1$) and times 
$t_i$ ($i=2\cdots N$), the shock velocity $v_s$, the dust radius parameters
$r_0$ and $t_0$, the $\tau_V=1$ radius of the progenitor wind $R(\tau_V=1)$,
and $Q_{rat}$, which enters into the estimated dust temperature and is a
proxy for the more complicated radiation transfer of the DUSTY models.

The emission has four components:  a shock break out 
of luminosity $L_{peak}$, duration $r_*/c$, and temperature $T_{peak}=50000$~K;  
an optical transient of luminosity $L_O= L(t)-L_N$; a shock with luminosity 
$L_S=L_N$; and, finally, the luminosity absorbed and re-radiated by the dust $L_D$.  
We model the
optical contributions as a black bodies, where it is useful to define  
\begin{equation}
      \hat{B}_\nu(n,T) = \lambda^{-n} B_\nu(T) \left[ \int d\nu \lambda^{-n} B_\nu(T) \right]^{-1},
\end{equation}
a modified Planck function normalized to unit total luminosity, and let
$f(\lambda) = \tau_\lambda/\tau_V$ be the optical depth ratio in the
DUSTY models between the visual optical depth $\tau_V$ and that at wavelength $\lambda$, $\tau_\lambda$.  The
escaping radiation from the transient is then 
\begin{equation}
      L_{\nu,O}(t) = L_0 \hat{B}_\nu(0,T(t)) e^{-\tau_V(t) f(\lambda)}
\end{equation}
and the fraction of the optical transient luminosity escaping is
\begin{equation}
      f_{O,esc} = \int d\nu B_\nu(0,T(t)) e^{-\tau_V(t) f(\lambda) }
\end{equation}
The shock luminosity is assumed to be initially generated as X-rays
at the energy of Eqn.~\ref{eqn:eshock}, and then absorbed by the wind
exterior to the shock using the density profile set by the progenitor
optical depth, $\rho = R(\tau_V=1)/\kappa r^2$, and the X-ray 
opacities of Eqn.~\ref{eqn:kapx}.  The escape fractions were 
computed for fixed wind densities and velocities of $v_s=1100$ and $560$~km/s, respectively.  
Let $f_{X,abs}$ be the fraction
of the X-rays energy which is absorbed.  We also apply the correction from Eqn.~\ref{eqn:tcool}
for the balance between cooling and expansion losses.  Thus, the
shock energy absorbed in the wind is
\begin{equation}
        L_{X,abs} = L_S f_{X,abs} \left[ 1 + t_{exp}/t_{cool} \right]^{-1}.
\end{equation}
In the simple model, we just treat this in the same manner is the radiation
from the optical transient, where it contributes
\begin{equation}
      L_{\nu,X}(t) = L_{X,abs} \hat{B}_\nu(0,T(t)) e^{-\tau_V(t) \tau_R(\lambda)}
\end{equation}
by direct emission and with the same fraction absorbed by dust.  The
shock model discussed in \S\ref{sec:optir} uses a list of emission lines and the
fraction of energy radiated in each line from \cite{Allen2008} as the
model for the spectrum produced by the absorbed X-rays, and computes
the extinction and fraction absorbed based on the model line spectrum.
Finally, the shock breakout luminosity was set by
\begin{equation} 
          L_{peak} = 16 \pi r_0^2 Q_{rat} \sigma T_{dest}^4
\end{equation}
with $T_{dest}=1500$~K, heating the dust at radius $r_0$ to $T_{dest}$.
We used a fixed temperature of $T_{peak}=50000$~K and gave it a duration
comparable to the light crossing time of the progenitor $r_*/c$ with
$r_* = 5 \times 10^{13}$~cm.  We only consider the ``dust echo''
from the breakout peak and ignore the short UV flash. 
  
The dust luminosity is then
\begin{equation}
        L_d(t) = (L_{peak}(t) + L_0(t) + L_{X,abs}(t)) (1-f_{O,esc})
\end{equation}
and the dust temperature is
\begin{equation}
        T_d = \left( { L_{peak}(t) + L_0(t) + L_{X,abs}(t)  \over 16 \pi \sigma Q_{rat} r_d(t)^2 }\right)^{1/4},
\end{equation}
including the factor $Q_{rat}$ (average ratios of Planck factors) to make it
easier to match this model to the DUSTY results.  We then used a modified
Planck function for the SED of the dust emission, 
\begin{equation}
        L_{\nu,d} = L_d \hat{B}_\nu(1,T_d).
\end{equation}
We then include the effects of light travel time (``dust echoes'') on the dust
emissions, spreading the contribution of the emission from any time $t$ 
uniformly over the interval $t < t' < t + 2 r(t)/c$, so the total emission is 
\begin{equation}
        L_\nu = L_{\nu,O}(t) + L_{\nu,X}(t) + \langle L_{\nu,d} (t') \rangle.
\end{equation}
We also experimented with including direct emission from the progenitor, modeled
as a black body with the luminosities from \S\ref{sec:progenitors} and a temperature
of $3000$~K.
The models were then fit to the light curve data, with priors to match the parameters
of the DUSTY models for $R(\tau_V=1)$ in the progenitor models and the sequence of 
luminosity, optical depth and dust radius estimates in the transient models. The
shock velocity was constrained to match the estimates by \cite{Smith2011} to 10\%
and the ``Planck'' term was constrained to have $\log Q_{rat}=0$ to 10\%.

\end{document}